\documentclass[aps,prd,12pt,preprint,superscriptaddress,preprintnumbers,nofootinbib,floatfix,longbibliography]{revtex4-1}

\usepackage[T1]{fontenc}
\usepackage[utf8]{inputenc}
\usepackage[english]{babel}
\usepackage{amsmath,amsthm,amssymb,amsfonts,mathrsfs,amsbsy,bm}
\usepackage{comment}
\usepackage{tensor}
\usepackage{slashed}
\usepackage{float}
 \usepackage{mathrsfs,tikz} 
\usepackage{esint}
\usepackage[a4paper, margin=1.5cm]{geometry}
\usepackage{cancel}
\usepackage{graphicx}
\usepackage{multirow}
\usepackage{array}
\usepackage{booktabs}
\usepackage{makecell}
\usepackage{xcolor}
\colorlet{BLUE}{blue}
\usepackage{tikz}
\usetikzlibrary{quotes,angles,arrows,decorations.markings,decorations.pathmorphing}
\usepackage{hyperref}
\hypersetup{colorlinks=true,breaklinks=true,citecolor=blue,linkcolor=[rgb]{0,0.5,0.9},urlcolor=blue}
\graphicspath{{figures/}}

\newcommand{\be}{\begin{equation}}
\newcommand{\ee}{\end{equation}}
\newcommand{\Be}{\begin{eqnarray}}
\newcommand{\Ee}{\end{eqnarray}}

\newcommand{\mincir}{\raise-3.truept\hbox{\rlap{\hbox{$\sim$}}\raise4.truept\hbox{$<$}\ }}
\newcommand{\magcir}{\raise-3.truept\hbox{\rlap{\hbox{$\sim$}}\raise4.truept\hbox{$>$}\ }}

\providecommand{\U}[1]{}

\newcommand{\ie}{\begin{equation}}
\newcommand{\fe}{\end{equation}}
\newcommand{\se}{\begin{eqnarray}}
\newcommand{\ff}{\end{eqnarray}}

\begin{document}
\emergencystretch=4em
\usetikzlibrary{arrows.meta,patterns}


\title{Accretion disks and photon rings in Palatini Kalb–Ramond gravity}

\author{A. A. Ara\'{u}jo Filho}
\email{dilto@fisica.ufc.br}
\affiliation{Departamento de F\'isica, Universidade Federal da Para\'iba, Caixa Postal 5008, 58051--970, Jo\~ao Pessoa, Para\'iba, Brazil.}
\affiliation{Departamento de F\'isica, Universidade Federal de Campina Grande, Caixa Postal 10071, 58429--900 Campina Grande, Para\'iba, Brazil.}
\affiliation{Center for Theoretical Physics, Khazar University, 41 Mehseti Street, Baku, AZ-1096, Azerbaijan.}

\author{N. Heidari}
\email{heidari.n@gmail.com}
\affiliation{Departamento de Física, Universidade Federal de Campina Grande, Caixa Postal 10071, 58429-900 Campina Grande, Paraíba, Brazil.}
\affiliation{Center for Theoretical Physics, Khazar University, 41 Mehseti Street, Baku, AZ-1096, Azerbaijan.}

\author{P. H. M. Barros}
\email{phmbarros@ufpi.edu.br}
\affiliation{Departamento de Matemática e F\'{i}sica, Universidade  Estadual do Maranhão (UEMA), Caxias - MA, 65604-380, Brazil}

 \author{Carlos F. S. Pereira} 
	\email{carlosfisica32@gmail.com}
	\affiliation{Departamento de Física, Universidade Federal do Espírito Santo, Avenida Fernando Ferrari, 514, Goiabeiras, 29060-900, Vit\'oria, Espir\'ito Santo, Brazil.}


\date{\today}


\begin{abstract}

We investigate thin accretion disks and photon rings around a static black hole in Palatini Kalb Ramond gravity, connecting thermal emission with the optical effects of the radial geometry. At fixed orbital mass and accretion rate, the local radiative flux scales as $1/\sigma$ and the effective temperature as $1/\sigma^{1/4}$, where $\sigma=\sqrt{1-q/2}$ and $q$ is the effective curvature coupling. The innermost stable circular orbit and the critical impact parameter share their Schwarzschild values, together with an emitted efficiency of approximately $5.72\%$. Moreover, photon trajectories admit an exact angular rescaling that determines the strong deflection coefficient and the exponential accumulation of successive disk images. We construct images of an opaque thermal disk and a transparent emitting sheet, taking into account gravitational frequency shifts, orbital motion, and radial emissivity. Positive $q$ increases the local thermal emission and accelerates the accumulation of successive images around the common critical curve. Under the universal coupling assumptions associated with the quoted Solar System bounds, the fractional corrections are of order $10^{-11}$ in flux and $10^{-12}$ in temperature.

\end{abstract}

\maketitle

\tableofcontents


\section{Introduction}

Accretion converts the gravitational binding energy of matter into radiation whose distribution carries information about the surrounding geometry. The spherical flow models established the relation between gravitational capture and the thermodynamic properties of the gas \cite{Bondi1952,Michel1972}. For matter with angular momentum, the thin disk descriptions connected orbital motion and angular momentum transport to the emitted energy flux \cite{Shakura1973,Novikov1973,Page1974}.

The radiation received by a distant observer also carries the effects of gravitational frequency shifts, Doppler modulation, and light deflection. Cunningham developed the relativistic transfer description of disk spectra, while Luminet constructed images displaying the combined influence of orbital motion and lensing \cite{Cunningham1975,Luminet1979}. Subsequent calculations extended this approach to tests of alternative spacetime geometries \cite{Bambi2012}. The Event Horizon Telescope images of M87* and Sagittarius A* further connected the observed emission rings and central brightness depressions to calculations of plasma emission and photon propagation \cite{EHT2019M87,EHT2022SgrA}.

Photons passing close to unstable circular orbits produce narrow images whose accumulation near the critical curve is described by the strong deflection expansion \cite{Bozza2002}. Analyses of lensing rings and photon rings clarified this image hierarchy and its interferometric signatures \cite{Gralla2019,GrallaLupsasca2020,Johnson2020}. Calculations for boson stars, regular compact objects, and parametrized black holes also showed how radial emissivity and the extent of the emitting material shaped the bright annuli and central depression \cite{Rosa2022,Olmo2023,Olmo2025}. Thermal disk studies of boson stars connected local dissipation to relativistic line profiles \cite{Rosa2024}.

Kalb Ramond gravity introduces an antisymmetric tensor field whose original formulation arose from interactions between strings \cite{Kalb1974}. Curvature couplings and a symmetry breaking potential allow its vacuum expectation value to select preferred spacetime directions \cite{Kostelecky2004Gravity,Altschul2010Tensor}. The propagator and excitation spectrum around such a vacuum were examined \cite{Maluf2018PropagatorKR}. The gravitational description subsequently developed through static black hole solutions, their thermodynamics, and classical tests \cite{Lessa2020KR,Yang2023KR}, followed by charged configurations \cite{Duan2024ChargedKR}. Rotating geometries and solutions obtained to first order in rotation extended the investigation to the joint influence of spin and the tensor background on light deflection and shadow contours \cite{Kumar2020RotatingKR,Liu2025SlowRotatingKR}.

The optical and dynamical response of these black holes has received substantial attention. It was examined the influence of the antisymmetric tensor on shadows and quasinormal frequencies \cite{Araujo2024ShadowsKR}. Complementary perturbation calculations addressed quasinormal spectra, time evolution, and greybody transmission in static geometries \cite{Guo2024QNMGreybodyKR}, with recent extensions to slowly rotating backgrounds \cite{Deng2026SlowRotationQNMKR}. Charged configurations were also investigated through geodesics, scattering, evaporation, and quasinormal ringing \cite{Araujo2025ChargedKR}. In addition, calculations of particle creation connected bosonic and fermionic emission to Hawking temperature, transmission through the curvature potential, and black hole evaporation \cite{Araujo2025EvaporationKR}.

The propagation of matter and radiation supplied further tests of the tensor background. Studies of charged black hole lensing determined weak and strong deflection, relativistic image positions, magnifications, and time delays \cite{Araujo2025LensingKR}. The optical mechanical description of particle motion also related the metric to dispersion, group velocity, and the thermodynamic properties of particle ensembles \cite{Araujo2025MotionThermalKR}. Moreover, extensions based on non--commutative geometry examined the correlation between the tensor background, quantum emission, and scalar perturbations \cite{Araujo2025NoncommKR}. Their optical properties were subsequently investigated through photon trajectories, shadows, and gravitational lensing \cite{Araujo2026OpticalNoncommKR}.

Accretion studies brought the dynamics of the surrounding matter into this description. It was investigated particle currents and mass accretion rates for a collisionless Vlasov gas \cite{Zhang2025VlasovKR}; examined disk radiative flux, temperature, luminosity, and orbital oscillations around charged Kalb--Ramond black holes \cite{Jumaniyozov2025RadiativeKR}. Calculations with prescribed thin disk emission and plasma further related the observed intensity to the emitting region and photon transport \cite{Tan2025ThinDiskKR,Xu2025OpticalKR}. More recent images of geometrically thick flows included polarized radiation and the dependence on emission anisotropy and observer inclination \cite{Zeng2026ThickPolarizedKR}.

The Palatini formulation treats the metric and affine connection as independent variables, with the connection equations introducing an auxiliary geometry in several classes of theories \cite{Olmo2009Dynamics,Olmo2011Palatini,Delhom2021Bumblebee}. Within Kalb--Ramond gravity, it was obtained exact static spherical solutions and determined their conserved mass and Solar System constraints \cite{She2026PalatiniKR}. An exact rotating vacuum solution subsequently extended the construction to stationary axisymmetric configurations and their orbital and optical properties \cite{Araujo2026RotatingPalatiniKR}. For the static solution, it was addressed the conserved mass from the orbital mass and examined particle motion, optical observables, and wave propagation \cite{Pereira2026Fingerprints}. At fixed orbital mass, the circular binding energy, the innermost stable circular orbit, and the critical impact parameter shared their Schwarzschild values, while proper radial distances and accumulated photon angles depended on the coupling.

In this paper, we investigate thin disk emission and photon transport in the static Palatini Kalb--Ramond geometry with a vanishing cosmological constant. We compare configurations at a common orbital mass $\mu$ and take into account the physical radial measure in the conservation equations and the conical asymptotics in photon propagation. The resulting thermal flux and its integrated energy budget are examined together with the exact angular rescaling of photon trajectories. We then construct images of an opaque thermal disk and a transparent sheet with prescribed radial emission, including gravitational frequency shifts and orbital motion. In this manner, we examine how the coupling redistributes local emission and changes the image features around a common critical curve.


\section{The black hole solution in Palatini Kalb Ramond gravity}
\label{sec:geometry}

We consider the static solution of Kalb Ramond gravity obtained through independent variation of the physical metric $g_{\mu\nu}$ and the symmetric affine connection $\Gamma^{\alpha}{}_{\mu\nu}$, with vanishing cosmological constant \cite{She2026PalatiniKR}. The antisymmetric field $B_{\mu\nu}$ extends the original Kalb Ramond construction \cite{Kalb1974} through curvature couplings that permit spontaneous Lorentz symmetry breaking \cite{Altschul2010Tensor,Kostelecky2004Gravity}. In geometrized units, $G=c=1$, the action reads
\begin{equation}
S=\frac{1}{2\kappa}\int\!\sqrt{-g}\left[P^{\mu\nu}R_{\mu\nu}(\Gamma)-\frac{1}{6}H^{\alpha\mu\nu}H_{\alpha\mu\nu}-V(\mathcal X)\right]\mathrm{d}^{4}x+S_{\mathrm m}[g,\psi],\qquad \kappa=8\pi,
\label{eq:action}
\end{equation}
where $g$ is the determinant of $g_{\mu\nu}$, $R_{\mu\nu}(\Gamma)$ is the Ricci tensor of the independent connection, $\psi$ denotes ordinary matter fields, and
\begin{equation}
P^{\mu\nu}=\left(1+\xi _1 B^{\alpha\beta}B_{\alpha\beta}\right)g^{\mu\nu}+\xi _2 B^{\mu\alpha}B^{\nu}{}_{\alpha},
\end{equation}
where $ H_{\alpha\mu\nu}=\partial_{[\alpha}B_{\mu\nu]}$, and $\mathcal X=B^{\mu\nu}B_{\mu\nu}+b^2$. Here, brackets denote antisymmetrization, $b>0$ fixes the vacuum norm, $V$ is the symmetry breaking potential, and $\xi _1$ and $\xi _2$ are coupling constants. Indices are moved with $g_{\mu\nu}$. We take into account the electric vacuum $B^{\mu\nu}B_{\mu\nu}=-b^2$, for which $H_{\alpha\mu\nu}=0$ and $V=V'=0$, with $V'=\mathrm{d}V/\mathrm{d}\mathcal X$.

The connection equation introduces an auxiliary metric $h_{\mu\nu}$ through $\sqrt{-h}\,h^{\mu\nu}=\sqrt{-g}\,P^{\mu\nu}$, where $h$ is its determinant. In this manner, $\Gamma^{\alpha}{}_{\mu\nu}$ is the Levi Civita connection of $h_{\mu\nu}$, following the auxiliary geometry construction familiar from Palatini theories \cite{Olmo2009Dynamics,Delhom2021Bumblebee}. Define the dimensionless combinations
\begin{equation}
\ell _1=\xi _1b^2,\qquad \ell _2=\xi _2b^2,\qquad a=1-\ell _1>0,\qquad q=\frac{\ell _2}{a}<2,\qquad \sigma=\sqrt{1-\frac q2},\qquad \rho=a\sigma.
\label{eq:couplings}
\end{equation}
The physical solution and its auxiliary geometry are
\begin{equation}
\mathrm{d}s_g^2=-F(r)\mathrm{d}t^2+\frac{\sigma^2}{F(r)}\mathrm{d}r^2+r^2\mathrm{d}\Omega_2^2,\qquad F(r)=1-\frac{2\mu}{r},\qquad \mu=\frac{M}{\rho},\qquad \mathrm{d}\Omega_2^2=\mathrm{d}\vartheta^2+\sin^2\vartheta\,\mathrm{d}\varphi^2,
\label{eq:metric}
\end{equation}
\begin{equation}
\mathrm{d}s_h^2=a\left[-F(r)\mathrm{d}t^2+\frac{\sigma^2}{F(r)}\mathrm{d}r^2+\sigma^2r^2\mathrm{d}\Omega_2^2\right],\qquad B_{tr}=\frac{b\sigma}{\sqrt2}.
\label{eq:auxiliary}
\end{equation}
The coordinate $r$ is the areal radius, $t$ has unit normalization at infinity, $M>0$ is the conserved mass obtained from the full gravitational action, and $\mu$ is the orbital mass measured by $\Omega^2r^3=\mu$ \cite{Iyer1994Noether,She2026PalatiniKR,Pereira2026Fingerprints}. All comparisons below fix $\mu$.

The substitutions $\bar t=\sqrt a\,t$ and $\bar r=\sqrt a\,\sigma r$ put Eq.~\eqref{eq:auxiliary} in Schwarzschild form, with auxiliary mass $\sqrt a\,\sigma\mu$. Thus $R_{\mu\nu}(\Gamma)=0$. The curvature of the physical metric, however, follows directly from Eq.~\eqref{eq:metric} and gives
\begin{equation}
R(g)=\frac{2(1-\sigma^{-2})}{r^2},\qquad r_{\mathrm h}=2\mu.
\label{eq:curvature}
\end{equation}
The event horizon lies at $r_{\mathrm h}=2\mu$. At large $r$, proper radial distance grows as $\sigma r$, while spherical area equals $4\pi r^2$. The exterior has conical asymptotics. This observation fixes the interpretation of angles and distances used in the images. Ordinary matter and electromagnetic radiation couple minimally to $g_{\mu\nu}$; their propagation follows its Levi Civita geometry. The disk is treated as a test source on the vacuum background, an approximation appropriate when its gravitational contribution is small.


\section{Circular motion and the accretion of matter}
\label{sec:accretion}

Consider an equatorial fluid element with four velocity $u^\alpha=\mathrm{d}x^\alpha/\mathrm{d}\tau$, where $\tau$ is proper time. Its specific energy $\mathcal E=-u_t$, specific angular momentum $\mathcal L=u_\varphi$, and angular velocity $\Omega=u^\varphi/u^t$ follow from circular geodesic motion. Introducing $x=r/\mu$, we obtain
\begin{equation}
\mathcal E(x)=\frac{1-2/x}{\sqrt{1-3/x}},\qquad \frac{\mathcal L(x)}{\mu}=\frac{\sqrt x}{\sqrt{1-3/x}},\qquad \mu\Omega=x^{-3/2},\qquad u^t=(1-3/x)^{-1/2}.
\label{eq:circular}
\end{equation}
The energy and $u^t$ are dimensionless, $\mathcal L$ has dimensions of length, and $\Omega$ has dimensions of inverse length. Radial stability gives the innermost stable circular orbit, abbreviated ISCO,
\begin{equation}
r_{\mathrm{ISCO}}=6\mu,\qquad \mathcal E_{\mathrm{ISCO}}=\sqrt{\frac89},\qquad \mathcal L_{\mathrm{ISCO}}=2\sqrt3\,\mu.
\end{equation}
These orbital quantities depend on the temporal and angular components of the metric. The radial component enters the frequency of small radial oscillations measured in $t$,
\begin{equation}
\kappa_r^2=\frac{\mu}{\sigma^2r^3}\left(1-\frac{6\mu}{r}\right).
\end{equation}
In this sense, the location of marginal stability and the timescale of the approach to it probe different aspects of the geometry \cite{Pereira2026Fingerprints}.

For the accreting gas, we adopt a stationary, geometrically thin, optically thick disk with efficient local cooling and zero viscous torque at $r_{\mathrm{ISCO}}$, as in the relativistic thin disk description \cite{Shakura1973,Novikov1973,Page1974}. The angular motion follows Eq.~\eqref{eq:circular} to leading order in the disk thickness. Let $z$ measure proper height from the equatorial plane, $H_{\mathrm d}$ be the proper half thickness, $\varrho_0$ be rest mass density, and $\Sigma=\int_{-H_{\mathrm d}}^{H_{\mathrm d}}\varrho_0\,\mathrm{d}z$ be surface density. Conservation of rest mass yields the positive inward rate per unit normalized Killing time $t$
\begin{equation}
\dot M_0=-2\pi\sigma r\Sigma u^r,\qquad \frac{\mathrm{d}\dot M_0}{\mathrm{d}r}=0.
\label{eq:massrate}
\end{equation}
The determinant entering the vertically integrated conservation equations is $\sqrt{-G}=\sigma r$ in coordinates $(t,r,z,\varphi)$; this defines $G$ and the radial measure appropriate to a thin disk.

The flux $\mathcal F(r)$ emitted from each face, measured per unit proper area and proper time in the fluid frame, is \cite{Page1974}
\begin{equation}
\mathcal F(r)=-\frac{\dot M_0}{4\pi\sigma r}\frac{\Omega_{,r}}{(\mathcal E-\Omega\mathcal L)^2}\int_{6\mu}^{r}(\mathcal E-\Omega\mathcal L)\mathcal L_{,\tilde r}\,\mathrm{d}\tilde r,
\label{eq:flux}
\end{equation}
where a comma denotes differentiation and $\tilde r$ is the integration variable. Besides its numerical evaluation, Eq.~\eqref{eq:flux} admits an elementary expression useful for examining the radial dependence. Let us define
\begin{equation}
\mathcal J(x)=\sqrt x-\sqrt6-\frac{\sqrt3}{2}\ln\!\left[\frac{\sqrt x-\sqrt3}{\sqrt x+\sqrt3}\frac{\sqrt6+\sqrt3}{\sqrt6-\sqrt3}\right],\qquad \widehat{\mathcal F}(x)=\frac{4\pi\mu^2\mathcal F(x)}{\dot M_0}.
\end{equation}
Direct integration gives
\begin{equation}
\widehat{\mathcal F}(x)=\frac{3\mathcal J(x)}{2\sigma x^{5/2}(x-3)},\qquad x\geq6,\qquad \widehat{\mathcal F}(x)\sim\frac{3}{2\sigma x^3}\quad(x\rightarrow\infty).
\label{eq:analyticflux}
\end{equation}
This expression sets the flux to zero at the inner edge and recovers the expected inverse cubic decay at large radius. At fixed $\mu$ and $\dot M_0$, its entire radial profile acquires the factor $\sigma^{-1}$.

\begin{figure}[h]
\centering
\includegraphics[width=0.485\linewidth]{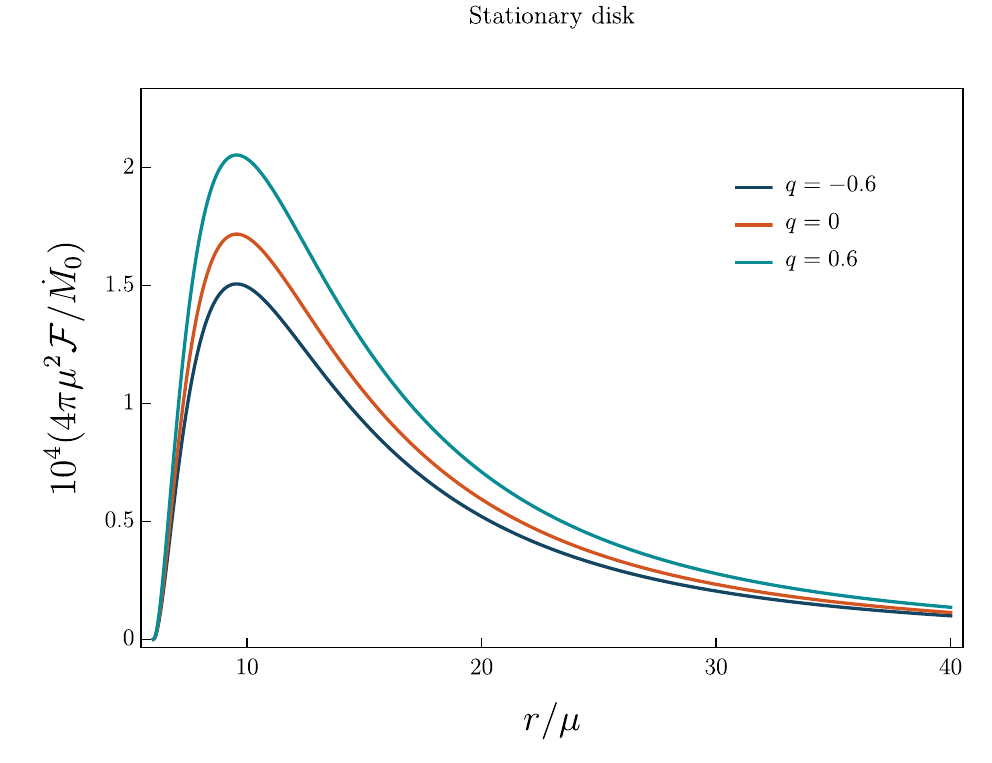}\hfill\includegraphics[width=0.485\linewidth]{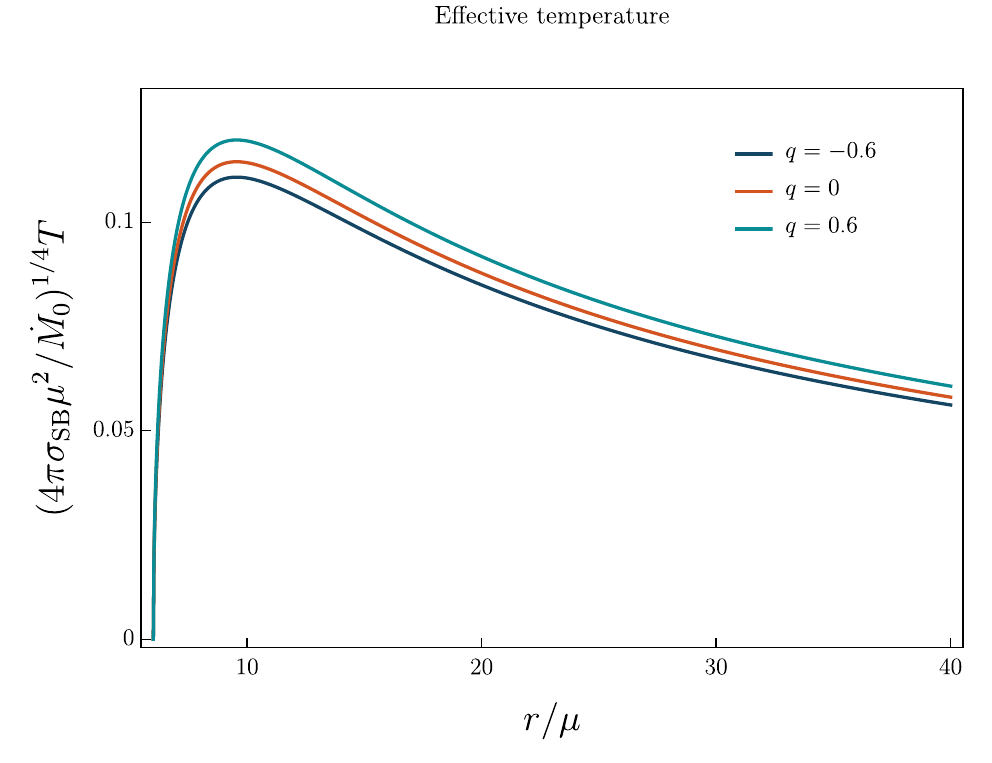}
\caption{Local radiative flux and effective temperature of the stationary disk for $q=-0.6,0,0.6$, at fixed orbital mass $\mu$ and rest mass supply $\dot M_0$. }
\label{fig:thermal}
\end{figure}

The effective temperature $T_{\mathrm{eff}}$ is defined by the Stefan Boltzmann law. With $\sigma_{\mathrm{SB}}$ denoting the Stefan Boltzmann constant, let
\begin{equation}
T_*=\left(\frac{\dot M_0}{4\pi\mu^2\sigma_{\mathrm{SB}}}\right)^{1/4},\qquad \widehat T(x)=\frac{T_{\mathrm{eff}}(x)}{T_*}=\widehat{\mathcal F}(x)^{1/4}.
\label{eq:temperature}
\end{equation}
Figure~\ref{fig:thermal} shows the resulting profiles. The maximum occurs at $x\simeq9.55093$ for every displayed coupling. For $q=0.6$, the local flux increases by approximately $19.5\%$ and the temperature by $4.56\%$ relative to $q=0$. For $q=-0.6$, the corresponding factors are $0.87706$ and $0.96774$. The position of the hottest annulus shows the common orbital scale, while its local emission responds to the proper radial geometry.

The energy budget takes into account the same radial measure. The rate of emitted Killing energy, integrated over both faces, is
\begin{equation}
L_{\mathrm{em}}=4\pi\int_{6\mu}^{\infty}\sigma r\mathcal E(r)\mathcal F(r)\,\mathrm{d}r=\dot M_0\left(1-\sqrt{\frac89}\right),\qquad \eta_{\mathrm{em}}=\frac{L_{\mathrm{em}}}{\dot M_0}\simeq0.057191.
\label{eq:luminosity}
\end{equation}
The factor $\sigma$ in the measure balances the local enhancement of $\mathcal F$. In other words, the redistribution per unit proper area is compatible with the same available orbital binding energy. Equation~\eqref{eq:luminosity} describes emitted energy under the thin disk assumptions. Photon capture and radiation returning to the surface determine the fraction that ultimately escapes, as established in relativistic disk transfer calculations \cite{Cunningham1975,Thorne1974}.

\section{Photon propagation and disk transfer functions}
\label{sec:photons}

Let us consider $k^\alpha=\mathrm{d}x^\alpha/\mathrm{d}\lambda$ be the photon tangent, with $\lambda$ an affine parameter. Spherical symmetry places each trajectory in a plane through the center. Its conserved energy is $E_\gamma=-k_t$, its total angular momentum is $L_\gamma$, and $b_\gamma=L_\gamma/E_\gamma$ is the impact parameter. Writing $\beta=b_\gamma/\mu$, $u=\mu/r$, and $\chi$ for the angular coordinate in that plane, the null equations become
\begin{equation}
\sigma^2\left(\frac{\mathrm{d}r}{\mathrm{d}\lambda}\right)^2=E_\gamma^2\left[1-\frac{b_\gamma^2F(r)}{r^2}\right],\qquad \left(\frac{\mathrm{d}u}{\mathrm{d}\chi}\right)^2=\frac{\beta^{-2}-u^2+2u^3}{\sigma^2},\qquad \frac{\mathrm{d}^2u}{\mathrm{d}\chi^2}=\frac{-u+3u^2}{\sigma^2}.
\label{eq:null}
\end{equation}
The unstable circular orbit and its critical impact parameter are
\begin{equation}
r_{\mathrm{ph}}=3\mu,\qquad b_{\mathrm c}=3\sqrt3\,\mu,\qquad \beta_{\mathrm c}=3\sqrt3.
\label{eq:critical}
\end{equation}
At a finite observer radius $x_{\mathrm o}=r_{\mathrm o}/\mu$, the inward initial data are $\chi=0$, $u=1/x_{\mathrm o}$, and $\mathrm{d}u/\mathrm{d}\chi>0$. For equal initial radius and impact parameter, the complete solution obeys the exact relation $u_q(\chi;\beta)=u_0(\chi/\sigma;\beta)$, where the subscript $0$ denotes $q=0$. The accumulated angle changes by $\sigma$, although the capture threshold remains at the common value in Eq.~\eqref{eq:critical}.

Equation~\eqref{eq:null} also admits an elliptic representation,
\begin{equation}
u_q(\chi;\beta)=\frac16+2\wp\!\left(\frac{\chi}{\sigma}+z_0;g_2,g_3\right),\qquad g_2=\frac1{12},\qquad g_3=\frac1{216}-\frac1{4\beta^2}.
\label{eq:elliptic}
\end{equation}
Here, $\wp$ is the Weierstrass elliptic function with invariants $g_2$ and $g_3$. The dimensionless phase $z_0$ is fixed by the observer initial data and the inward radial orientation. The critical trajectory follows from the smooth limiting form. This expression gives an independent description of the highly wound rays and resolves the short impact intervals occupied by successive images.

\begin{figure}[h]
\centering
\includegraphics[width=0.485\linewidth]{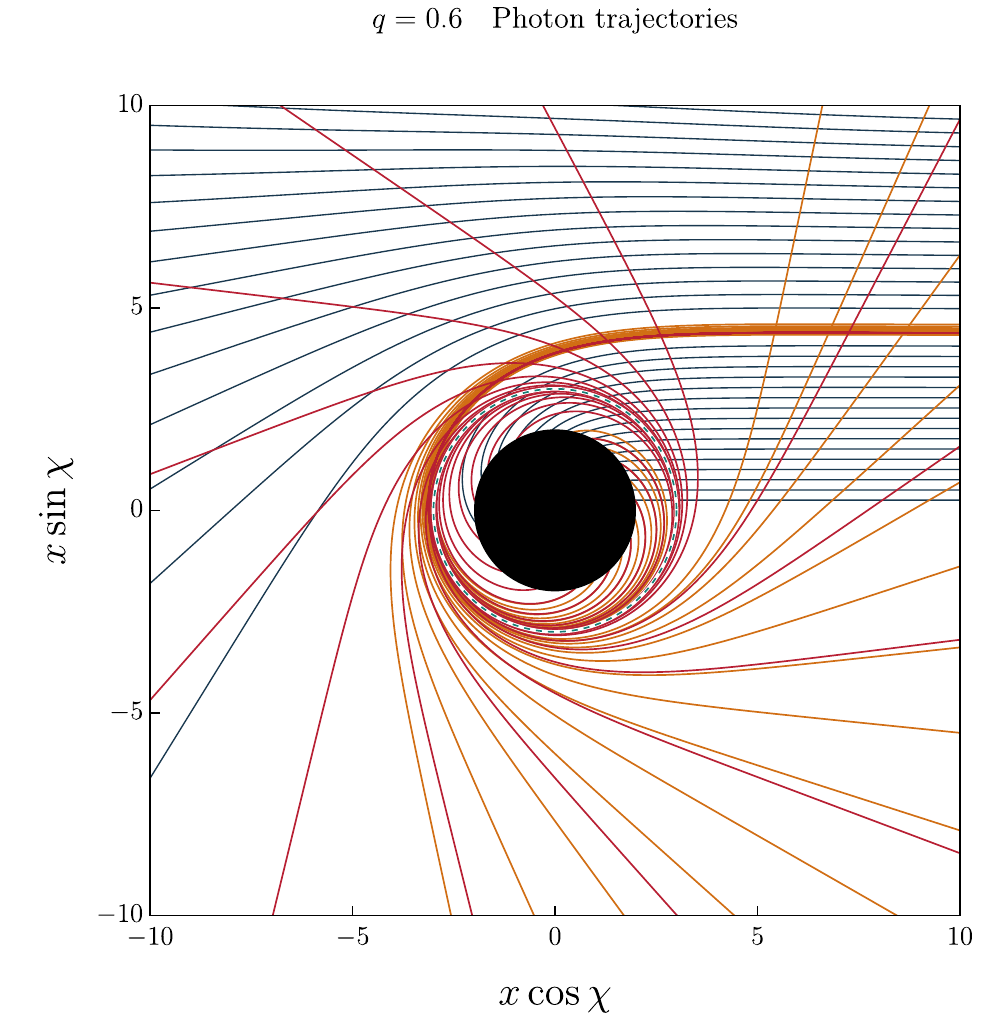}\hfill\includegraphics[width=0.485\linewidth]{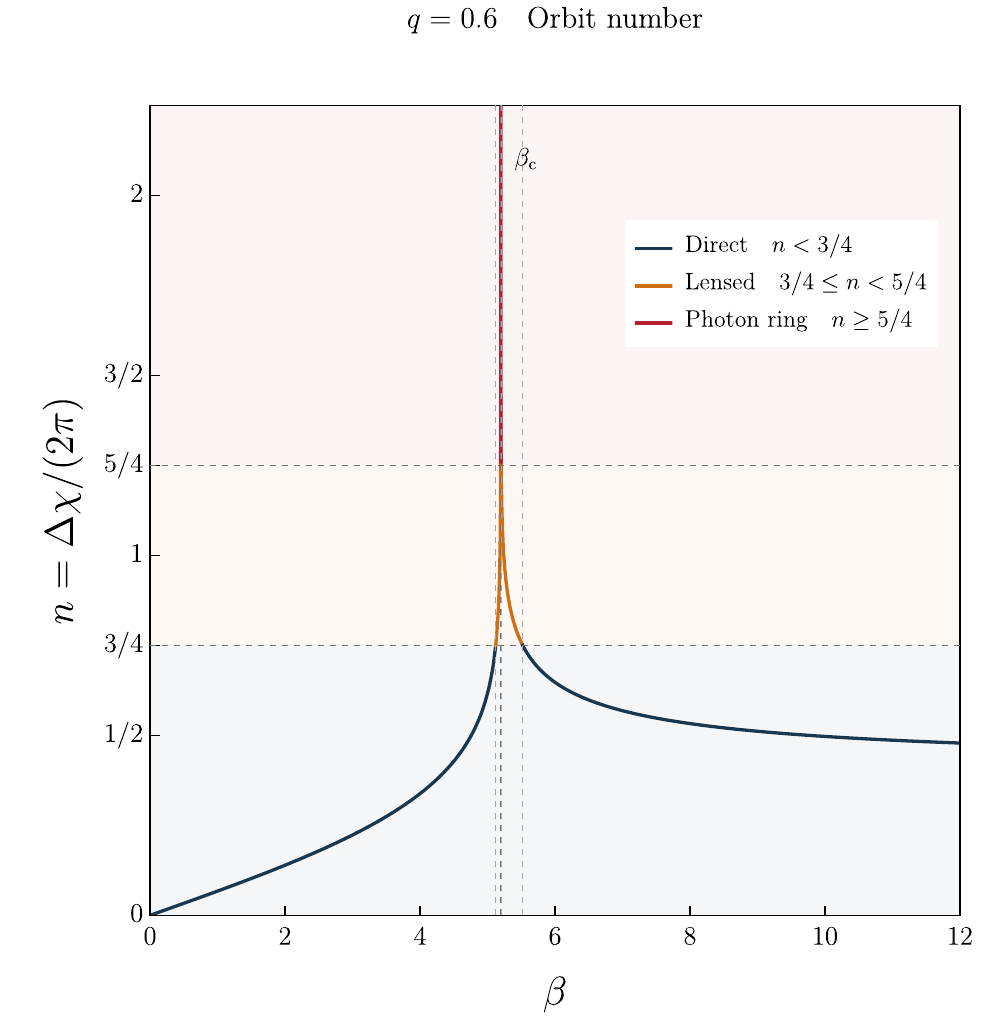}
\caption{Dense ray tracing and the corresponding orbit number at $q=0.6$. The $82$ trajectories share the winding colors of the right panel: direct rays in blue, lensed rays in orange, and the photon ring class in crimson. The black disk marks the horizon at $r=2\mu$, while the dashed circle marks the photon orbit at $r=3\mu$. Capture and escape occur within each winding class. The shaded bands and horizontal guides indicate the thresholds $n=3/4$ and $n=5/4$.}
\label{fig:orbits}
\end{figure}

For the orbit number calculation, $n=\Delta\chi/(2\pi)$ measures the total winding from infinity to the horizon for capture, and from infinity back to infinity for scattering. In the conventional axial classification, the intervals $n<3/4$, $3/4\leq n<5/4$, and $n\geq5/4$ identify direct, lensed, and highly wound trajectories \cite{Gralla2019}. Figure~\ref{fig:orbits} follows these classes across both capture and escape, with additional rays concentrated near the critical impact parameter. For $q=0.6$, the photon ring class occupies approximately $5.19435\leq\beta\leq5.20293$, and $\beta_{\mathrm c}=5.19615$ separates the two outcomes. The intersection index introduced below counts the actual encounters with the emitting material. This distinction becomes useful when comparing sources with different radial support \cite{Olmo2023,Rosa2022}.

The trajectory plane uses the polar projection $((r/\mu)\cos\chi,(r/\mu)\sin\chi)$. Incoming rays approach from the right with $(r/\mu)\sin\chi\rightarrow\sigma\beta$ at large radius. In this manner, their parallel asymptotes take into account the conical geometry, while $\beta$ keeps its definition through the conserved photon energy and angular momentum.

The strong deflection expansion gives rise to an analytic description of the narrow image hierarchy \cite{Bozza2002,GrallaLupsasca2020}. For scattering between asymptotic regions, let us regard that $\alpha=\Delta\chi-\pi$ and $\delta=\beta/\beta_{\mathrm c}-1>0$. In this regard, we obtain
\begin{equation}
\alpha_q=\sigma\alpha_0+\pi(\sigma-1)=-\sigma\ln\delta+\sigma\ln\!\left[216(7-4\sqrt3)\right]-\pi+O(\delta\ln\delta).
\label{eq:deflection}
\end{equation}
The logarithmic coefficient is $\bar a=\sigma$. The constant term $\pi(\sigma-1)$ in the first equality expresses the conical asymptotic geometry. Close to the critical curve, the radial displacement $\Delta\beta_m=|\beta_m-\beta_{\mathrm c}|$ of successive disk images sampling the same emitting annulus satisfies
\begin{equation}
\frac{\Delta\beta_{m+1}}{\Delta\beta_m}\longrightarrow\exp\!\left(-\frac{\pi}{\sigma}\right),\qquad \lambda_t=\frac{1}{3\sqrt3\,\sigma\mu}.
\label{eq:subrings}
\end{equation}
Here, $m$ labels successive disk crossings, and $\lambda_t$ is the radial instability exponent of the photon orbit measured in $t$. Equation~\eqref{eq:subrings} connects the angular accumulation with the exponential narrowing of successive images, as in photon ring analyses \cite{Johnson2020,GrallaLupsasca2020}. The narrowing factor equals $0.06359$, $0.04321$, and $0.02340$ for $q=-0.6,0,0.6$, respectively.

\begin{figure}[h]
\centering
\includegraphics[width=0.485\linewidth]{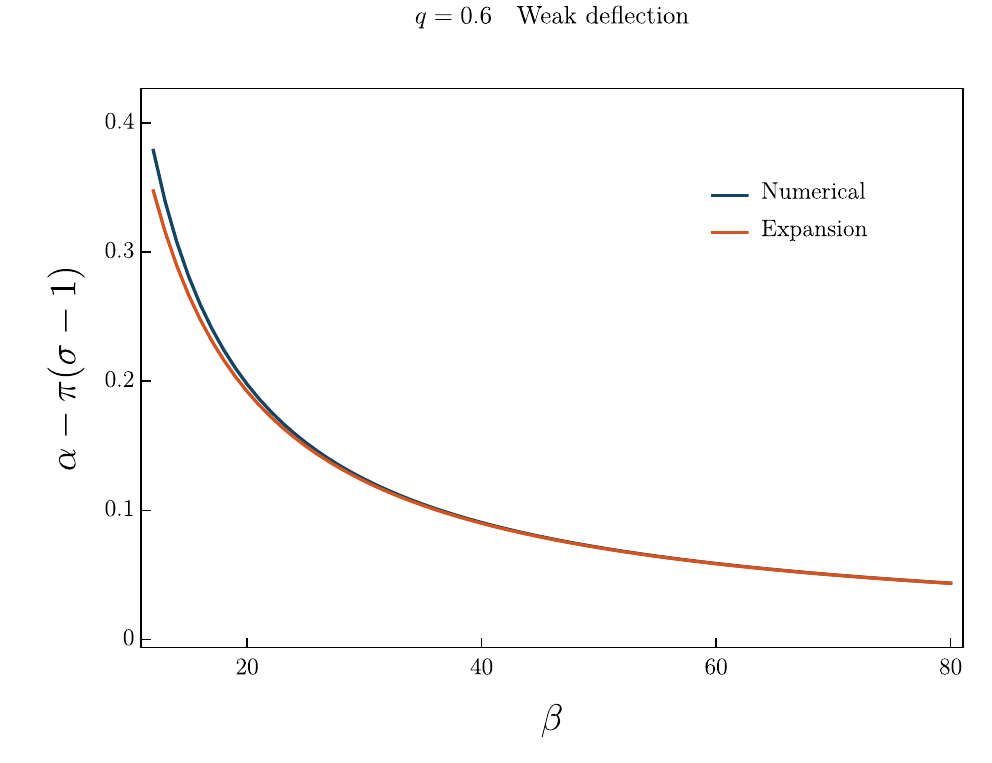}\hfill\includegraphics[width=0.485\linewidth]{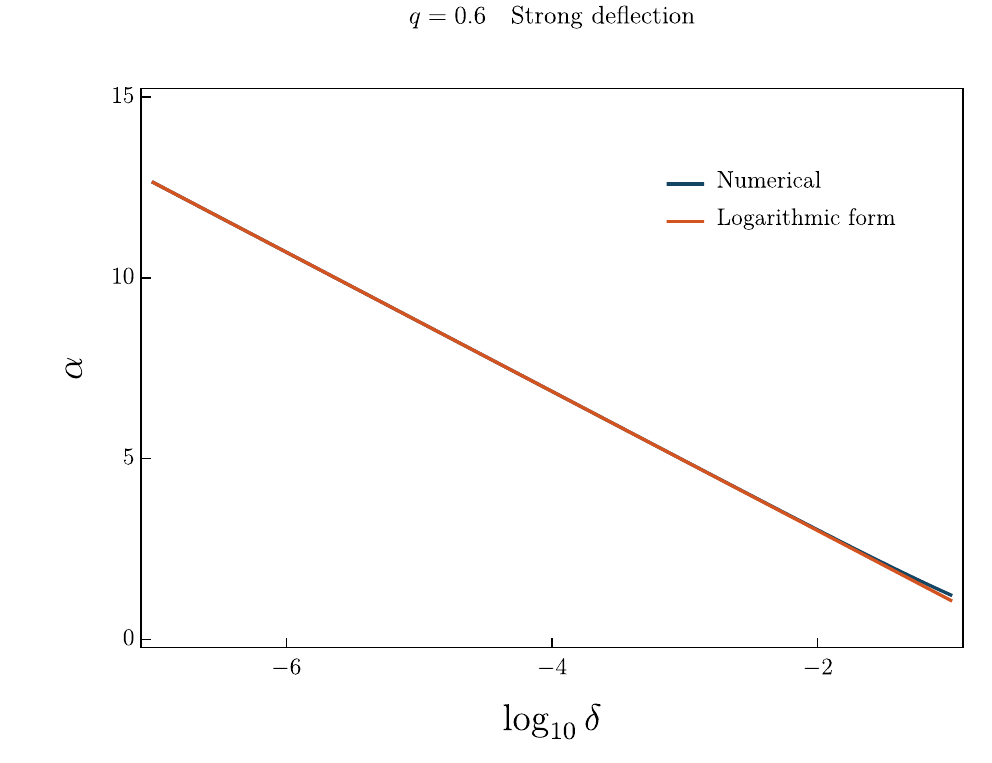}
\caption{Numerical deflection and its analytic limits at $q=0.6$. The left panel subtracts the conical contribution $\pi(\sigma-1)$ and compares the mass dependent bending with Eq.~\eqref{eq:weakdeflection}. The right panel uses $\delta=\beta/\beta_{\mathrm c}-1$ and compares the exact quadrature with the logarithmic expression in Eq.~\eqref{eq:deflection}.}
\label{fig:deflection}
\end{figure}

The weak deflection expansion separates the background angular contribution from the bending produced by the central mass. The exact angular rescaling gives
\begin{equation}
\alpha_q=\pi(\sigma-1)+\sigma\left[\frac{4}{\beta}+\frac{15\pi}{4\beta^2}+O(\beta^{-3})\right].
\label{eq:weakdeflection}
\end{equation}
Figure~\ref{fig:deflection} compares this expression and the strong deflection form with independent radial quadrature. The weak regime resolves the mass dependent contribution on the conical background, while the logarithmic regime describes the accumulation of rays close to the photon orbit. Together they connect the broad deflection of incoming radiation with the narrow image hierarchy \cite{Bozza2002,Pereira2026Fingerprints}.

\begin{figure}[h]
\centering
\includegraphics[width=0.485\linewidth]{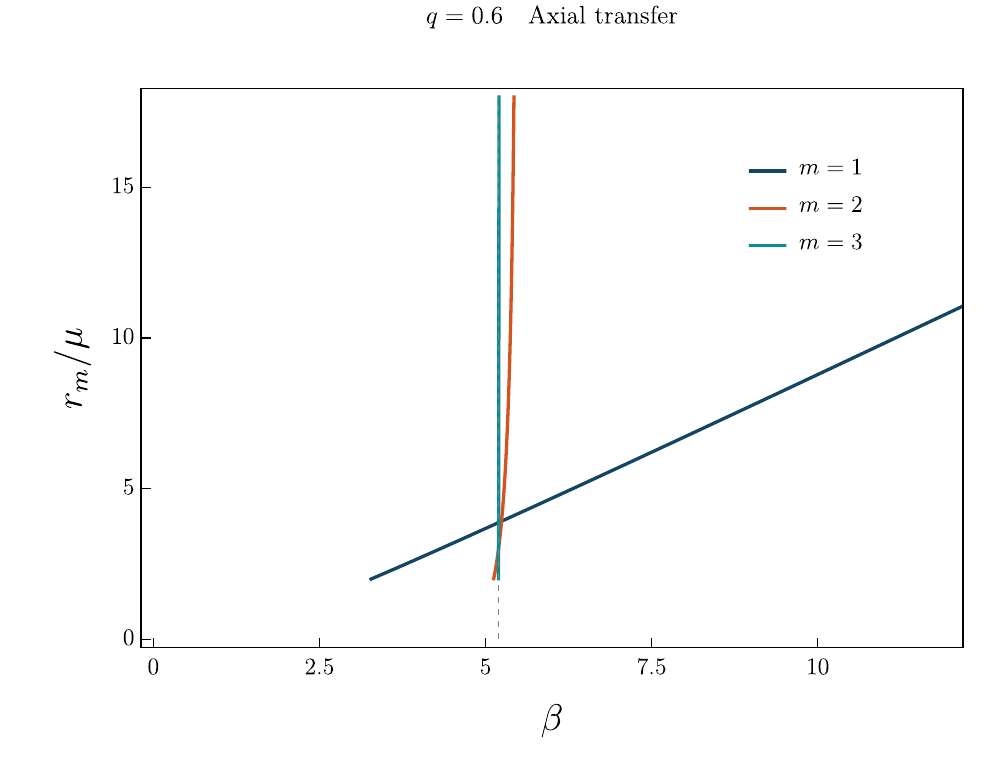}\hfill\includegraphics[width=0.485\linewidth]{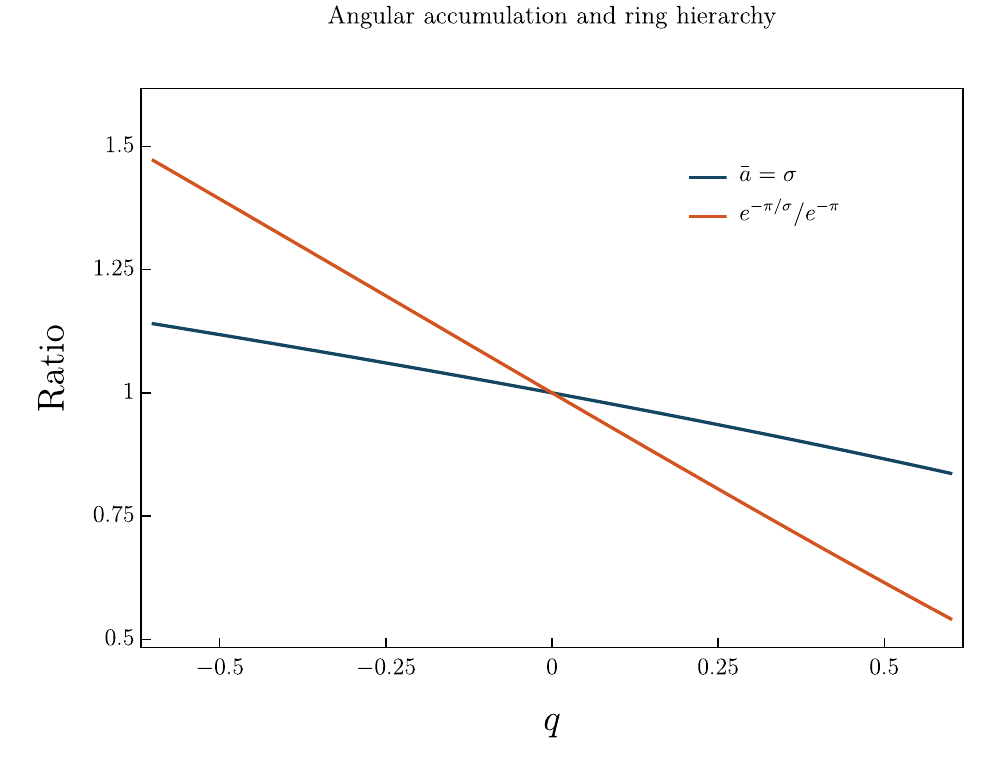}
\caption{Axial transfer functions and analytic lensing dependence. The transfer functions map screen impact parameter to successive emission radii at $q=0.6$. }
\label{fig:transfer}
\end{figure}

For imaging, place a static observer at $x_{\mathrm o}=r_{\mathrm o}/\mu=1000$ and inclination $i$ from the disk normal. The dimensionless impact coordinates $X$ and $Y$ on the observer sky obey $\beta^2=X^2+Y^2$ and $k_\varphi/(\mu E_\gamma)=-X\sin i$. The local angular separation $\alpha_{\mathrm{sky}}$ from the inward radial direction is
\begin{equation}
\sin\alpha_{\mathrm{sky}}=\frac{\beta\sqrt{F(r_{\mathrm o})}}{x_{\mathrm o}}.
\end{equation}
With $\chi=0$ at the observer, the intersections with the equatorial plane satisfy
\begin{equation}
\cos i\cos\chi+\frac{Y}{\beta}\sin i\sin\chi=0,\qquad \chi_m=\chi_1+(m-1)\pi,\qquad x_m(X,Y)=\frac{1}{u_q(\chi_m;\beta)}.
\label{eq:crossings}
\end{equation}
The first positive solution defines $\chi_1$, and $m=1,2,\ldots$ orders encounters along the backward ray. For an axial observer, $i=0$ and $\chi_m=(m-1/2)\pi$. Figure~\ref{fig:transfer} shows the corresponding transfer functions. The domain of each function consists of rays that reach that crossing before capture or escape. Furthermore, this construction follows the disk imaging methods developed for regular objects and parametrized geometries \cite{Olmo2023,Olmo2025}.


\section{Emission and the optical appearance of the disk}
\label{sec:emission}

The thermal disk of Sec.~\ref{sec:accretion} supplies an emission model tied to the rest mass flow. For an isotropic blackbody surface, its comoving bolometric intensity is $I_{\mathrm e}=\mathcal F/\pi$. At frequency $\nu$, the corresponding specific intensity is
\begin{equation}
B_\nu(T)=\frac{2h_{\mathrm P}\nu^3}{\exp[h_{\mathrm P}\nu/(k_{\mathrm B}T)]-1},
\end{equation}
where $T=T_{\mathrm{eff}}$, $h_{\mathrm P}$ is Planck's constant and $k_{\mathrm B}$ is Boltzmann's constant, with $c=1$ as above. A backward ray samples the first emitting surface that it reaches; crossings of the empty region inside $6\mu$ allow the ray to continue. This prescription describes an opaque disk and gives the usual connection between orbital dissipation and the measured spectrum \cite{Cunningham1975,Luminet1979,Rosa2024,Bambi2012}.

\begin{figure}[h]
\centering
\includegraphics[width=0.33\linewidth]{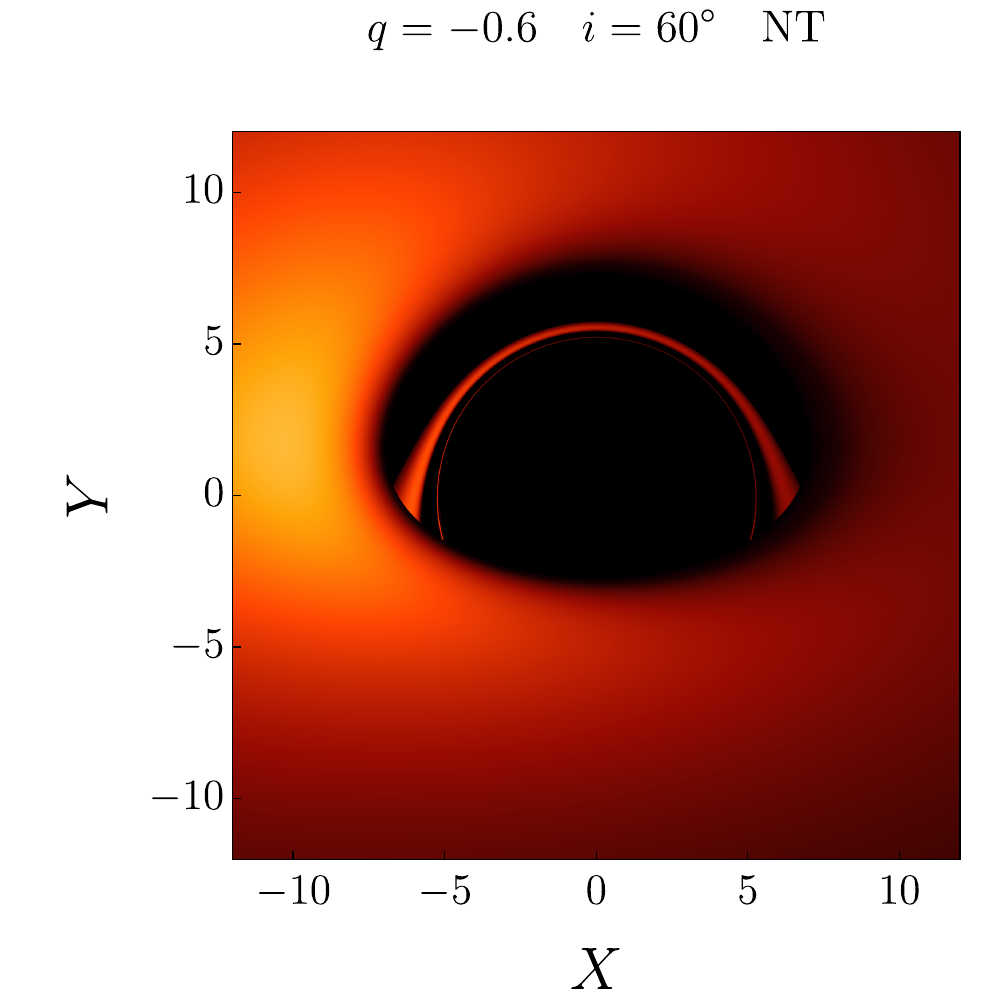}\includegraphics[width=0.33\linewidth]{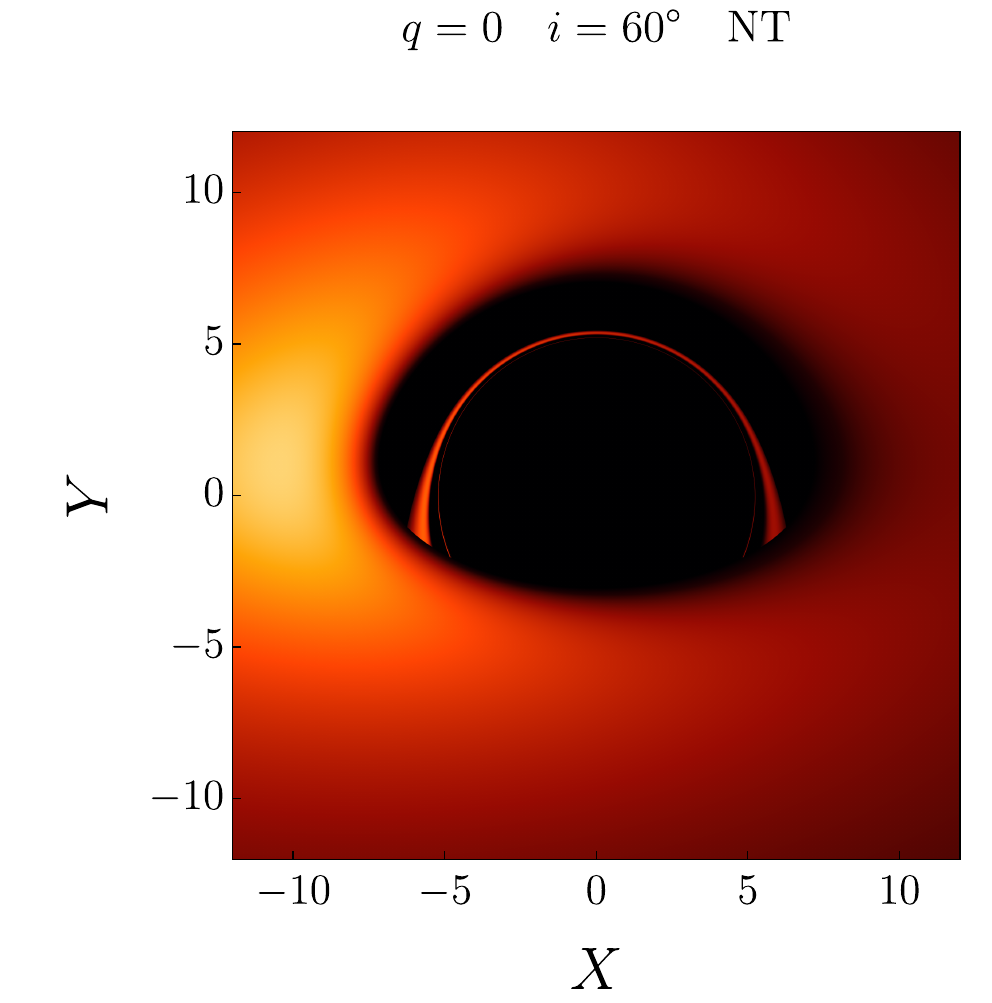}\includegraphics[width=0.33\linewidth]{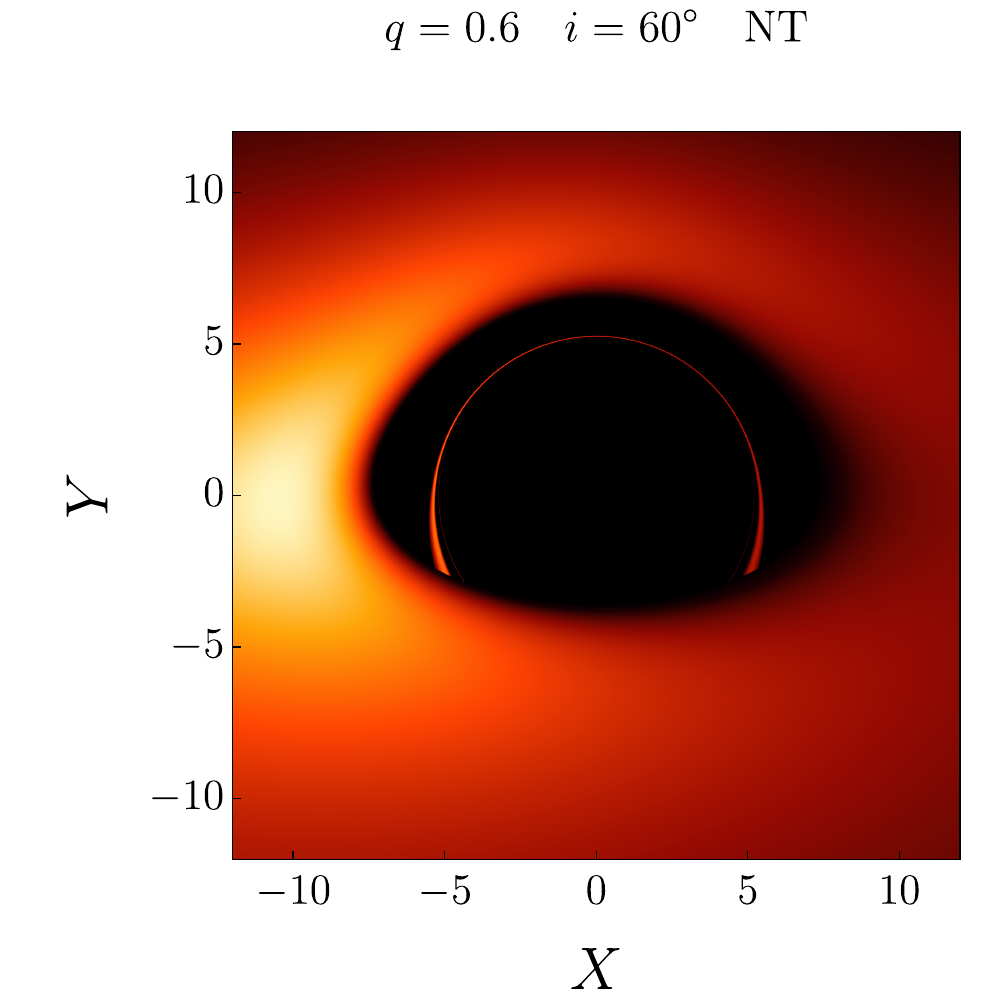}\includegraphics[height=0.27\linewidth]{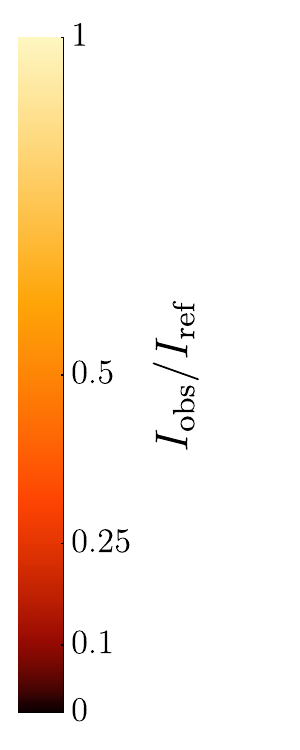}
\caption{Thermal images of the opaque accretion disk at $i=60^\circ$, with $q=-0.6,0,0.6$ from left to right. The flux follows Eq.~\eqref{eq:analyticflux} and the emitting surface extends over $6\leq x\leq40$. The panels share one intensity reference and square root display stretch. The color scale gives the linear ratio $I_{\mathrm{obs}}/I_{\mathrm{ref}}$.}
\label{fig:thermalimages}
\end{figure}

Figure~\ref{fig:thermalimages} combines local dissipation with photon transport. In each image comparison, $I_{\mathrm{ref}}$ is the largest sampled intensity in that group; axial comparisons use the resolved radial curves, and inclined comparisons use the camera maps. The display coordinate is $\sqrt{I_{\mathrm{obs}}/I_{\mathrm{ref}}}$. The thermal annuli have the same areal location across the three geometries, while their emitted intensity scales as $\sigma^{-1}$. Their observed brightness also takes into account the source radius reached by each ray and its orbital frequency shift. The luminous side of the disk thus carries both the accretion flux and the Doppler weighting, with the finer lensed emission recording the angular accumulation near the photon orbit.

To examine source morphology independently, we additionally use a transparent emitting sheet with prescribed surface intensity, following the optical studies  \cite{Rosa2022,Rosa2023,Olmo2023,Olmo2025}. In this prescription, each crossing supplies an isotropic intensity directly, and radiation propagates freely between crossings. The Johnson SU profile is
\begin{equation}
\begin{gathered}
J(x;\gamma,x_{\mathrm c},w)=\frac{\exp\!\left\{-\frac12\left[\gamma+\operatorname{arcsinh}\!\left(\frac{x-x_{\mathrm c}}{w}\right)\right]^2\right\}}{\sqrt{(x-x_{\mathrm c})^2+w^2}},\\ I_{\mathrm{em}}(x)\equiv I_{\mathrm e}(x)=\frac{I_0}{\mathcal N}J(x;\gamma,x_{\mathrm c},w)\,\mathbf1_{[x_{\mathrm{in}},40]}(x).
\end{gathered}
\label{eq:su}
\end{equation}
Here, $\gamma$ sets the asymmetry, $x_{\mathrm c}$ the radial location, $w>0$ the width, $I_0$ the common intensity scale, and $\mathbf1$ the indicator of the emitting interval. All three profile parameters are dimensionless. We take $(\gamma,x_{\mathrm c},w)=(-2,17/3,1/4)$ for the main SU2 comparison and define $\mathcal N$ as the maximum of $J$ on $2\leq x\leq40$. For supported static emitters, $x_{\mathrm{in}}=2$ is the limiting inner edge approached from the exterior; circular emitters have $x_{\mathrm{in}}=6$ and the same $\mathcal N$. The emission panel additionally displays the static SU1, SU5, and Horizon profiles, with parameter triplets $(-2,4,3/2)$, $(0,2,1)$, and $(-3,2,1/8)$, respectively, each normalized to its own maximum on $2\leq x\leq40$. The code includes the ten SU families explored in Ref.~\cite{Olmo2025}.

\begin{figure}[h]
\centering
\includegraphics[width=0.485\linewidth]{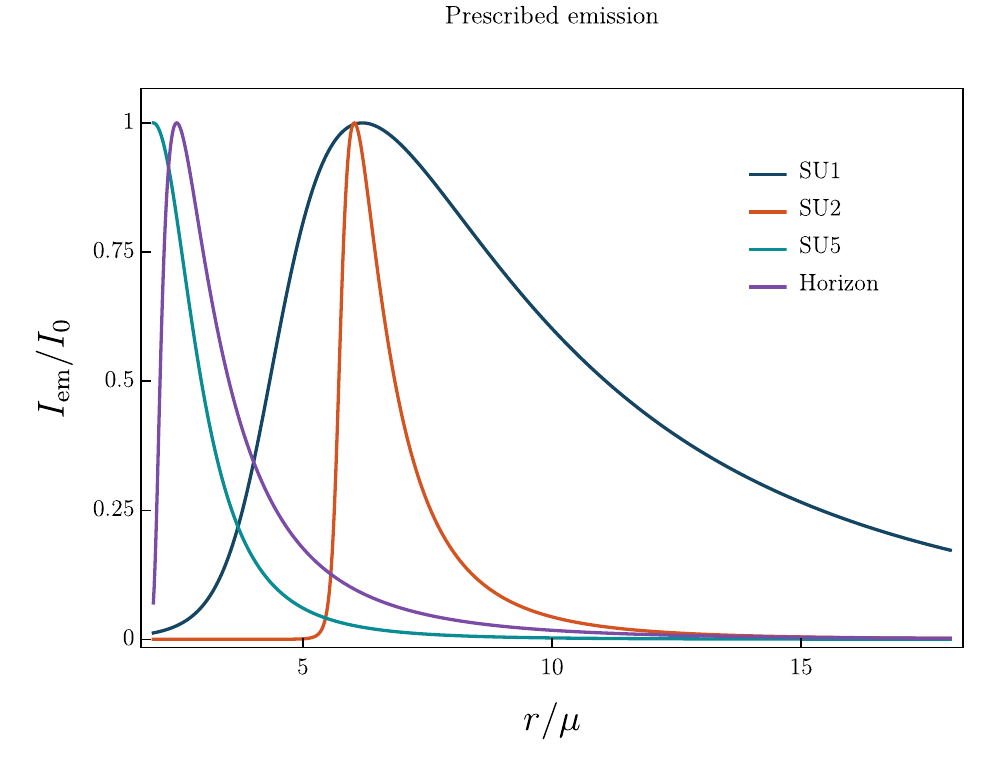}\hfill\includegraphics[width=0.485\linewidth]{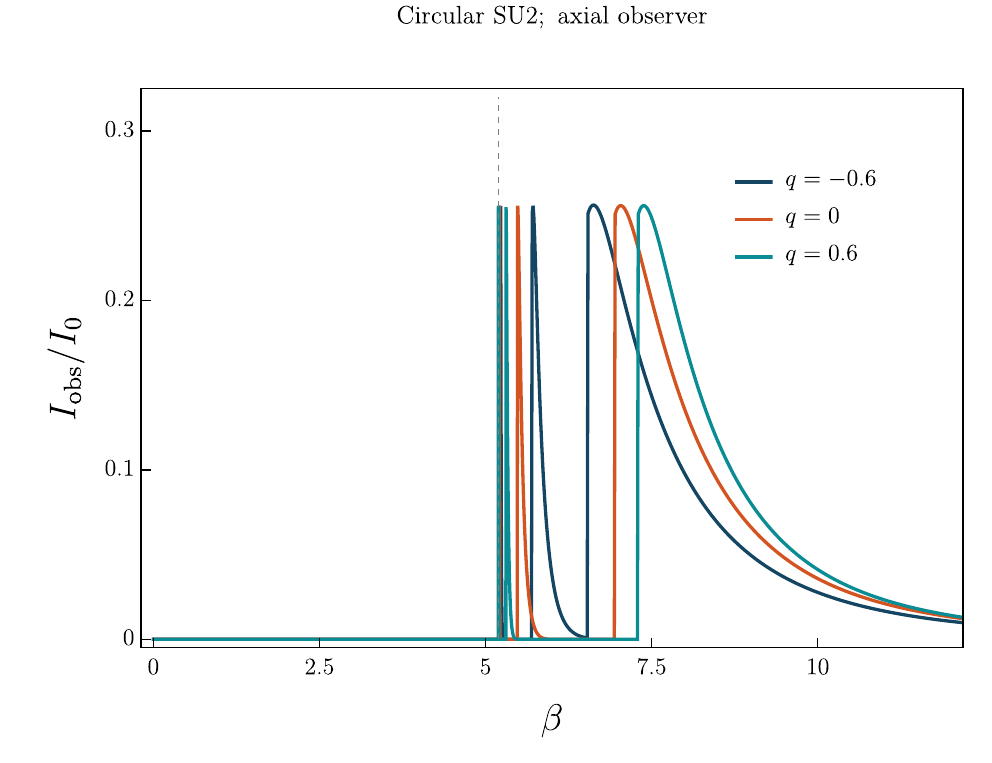}
\caption{Prescribed radial emission and the axial intensity of the transparent sheet. The source parameters and normalization are shared by the geometries being compared. The observed curves use circular emitters and include gravitational and orbital frequency shifts together with successive disk intersections. The critical curve remains at $\beta_{\mathrm c}=3\sqrt3$.}
\label{fig:intensity}
\end{figure}

Let us consider that $g_\nu=\nu_{\mathrm o}/\nu_{\mathrm e}$ be the ratio of observed to emitted frequency. Conservation of $I_\nu/\nu^3$ along a vacuum ray gives \cite{Cunningham1975,Dexter2009}
\begin{equation}
I_{\nu,\mathrm o}(\nu_{\mathrm o})=g_\nu^3 I_{\nu,\mathrm e}(\nu_{\mathrm o}/g_\nu),\qquad I_{\mathrm o}=g_\nu^4 I_{\mathrm e}.
\end{equation}
The bolometric intensity $I_{\mathrm o}$ is integrated over frequency. For the static and circular emitters, respectively, the shifts are
\begin{equation}
g_{\nu,\mathrm{stat}}=\sqrt{\frac{F(r)}{F(r_{\mathrm o})}},\qquad g_{\nu,\mathrm{circ}}=\frac{\sqrt{1-3/x}}{\sqrt{F(r_{\mathrm o})}\left[1+x^{-3/2}X\sin i\right]}.
\label{eq:shift}
\end{equation}
The second expression takes into account the gravitational shift, orbital time dilation, and the Doppler modulation. With this screen convention, the approaching side has $X<0$. At a fixed emission radius and screen position, these frequency factors have the same functional form for all $q$. The radial deformation enters through the emission radii sampled along the ray.

For the transparent sheet, the observed image is
\begin{equation}
I_{\mathrm{obs}}(X,Y)=\sum_{m=1}^{m_{\mathrm{max}}}g_{\nu,m}^4 I_{\mathrm e}(x_m),
\label{eq:image}
\end{equation}
where $m_{\mathrm{max}}=7$ is the numerical crossing cutoff. For the opaque thermal surface, the same transfer law is evaluated at the first crossing within its emitting support. All rendered images use an outer emitting radius $x_{\mathrm{out}}=40$; the analytic stationary disk and its energy budget extend to infinity. These two prescriptions allow the image response to be read with its appropriate physical emission assumptions. In particular, the freely specified $I_0$ of Eq.~\eqref{eq:su} characterizes the sheet, while $\dot M_0$ determines the thermal emission through Eq.~\eqref{eq:flux}.

\begin{figure}[h]
\centering
\includegraphics[width=0.35\linewidth]{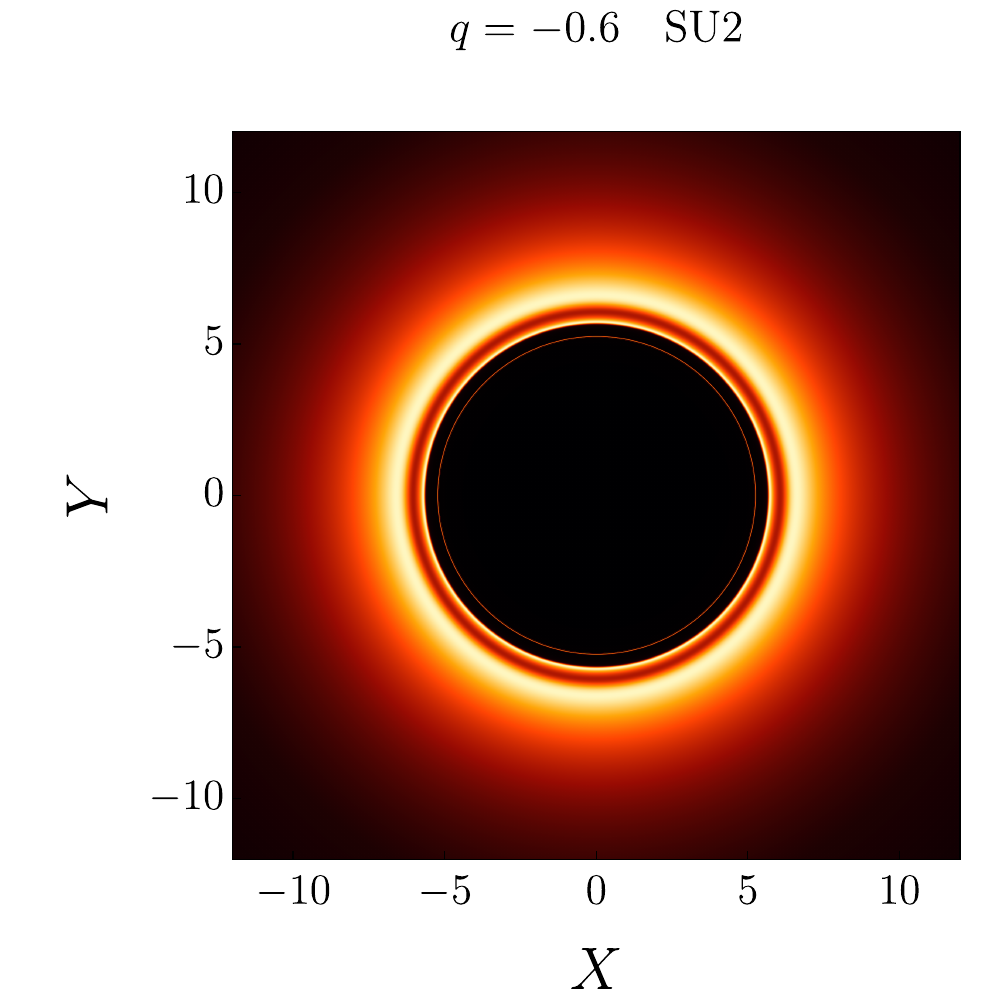}\includegraphics[width=0.35\linewidth]{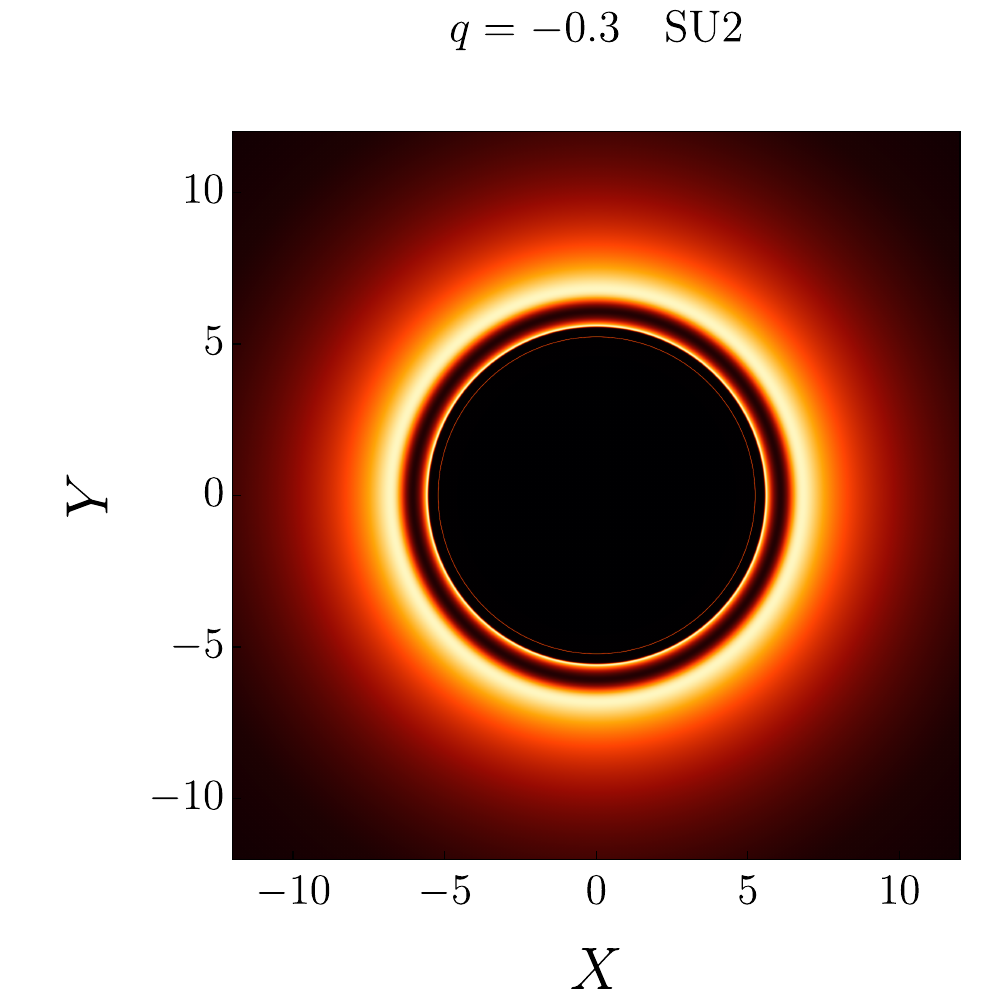}\par\vspace{1ex}
\includegraphics[width=0.35\linewidth]{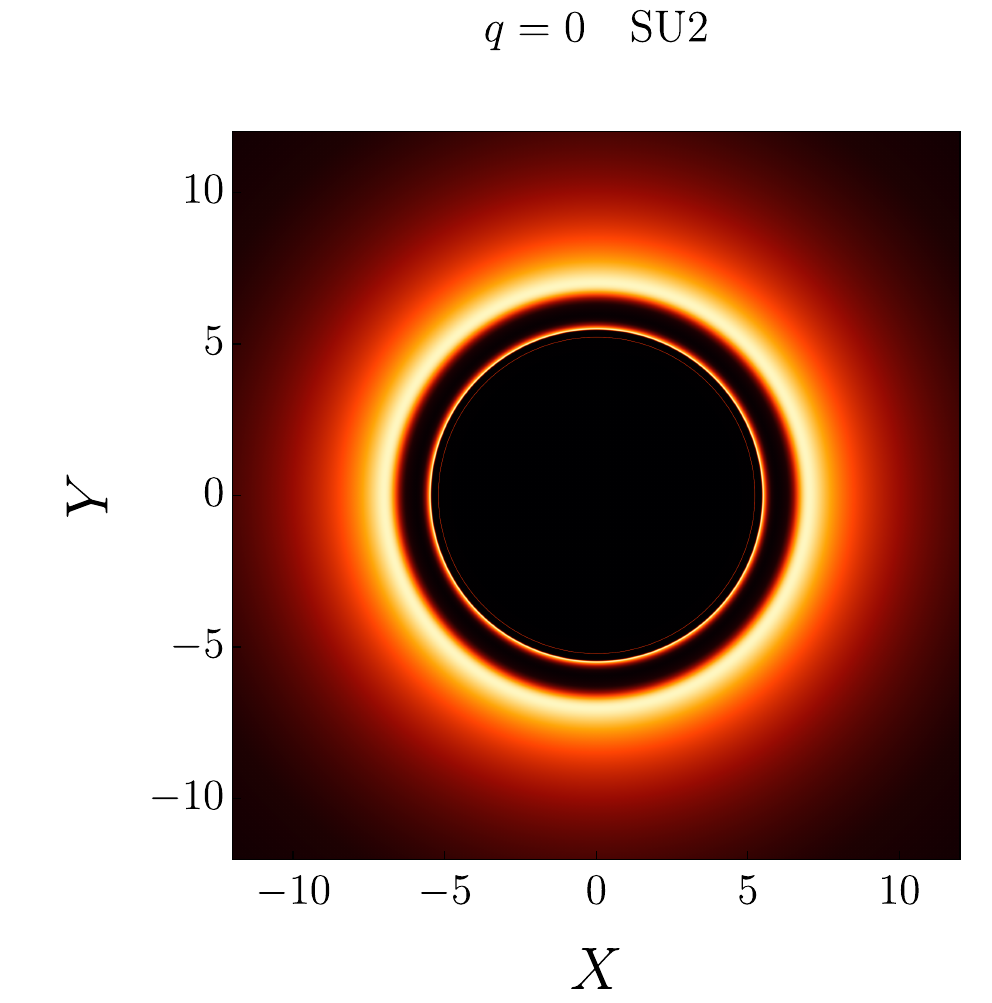}\includegraphics[width=0.35\linewidth]{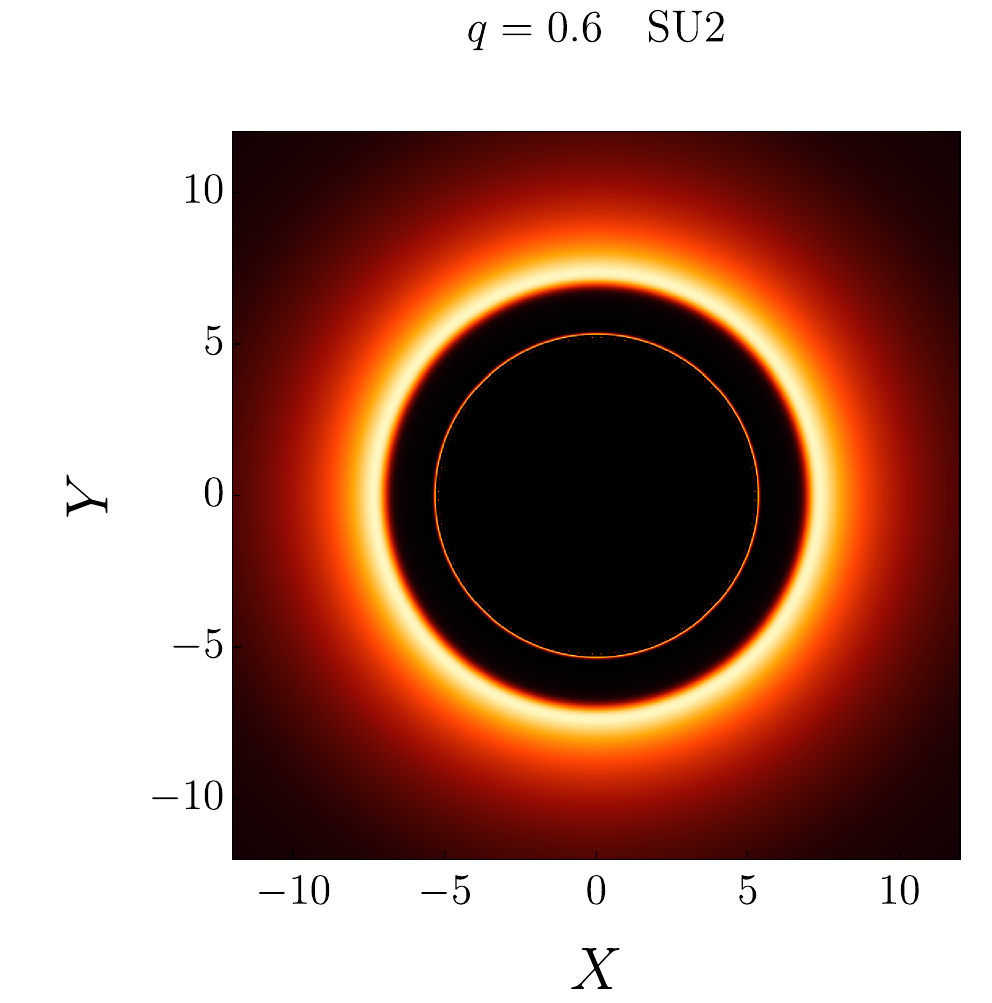}\par\vspace{1ex}
\includegraphics[width=0.5\linewidth]{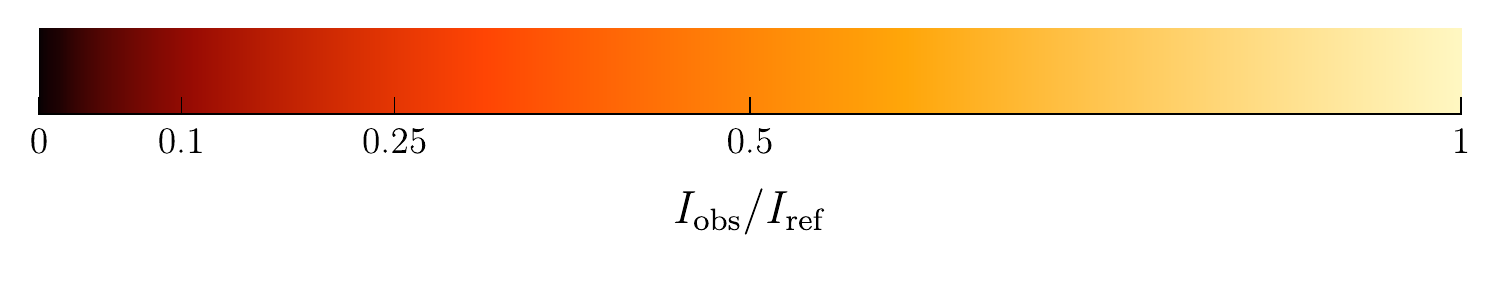}
\caption{Face on images of the fixed SU2 profile with static emitters. The couplings are $q=-0.6,-0.3,0,0.6$, ordered from the upper left to the lower right. All maps share the orbital mass, emission profile, source normalization, field of view, and intensity reference. The horizontal color scale gives $I_{\mathrm{obs}}/I_{\mathrm{ref}}$; the display applies the same square root stretch to every panel.}
\label{fig:axialmaps}
\end{figure}

For an axial observer, the static sheet gives a direct view of the radial emission and its successive lensed images. Figure~\ref{fig:axialmaps} compares the same SU2 source at $q=-0.6,-0.3,0,0.6$, with one brightness reference across all four maps. The broad direct annulus moves outward as $q$ increases, sampling the same radial emission profile through a changing transfer function. At a fixed emission radius, the gravitational weighting is common to these geometries. The isolated direct maximum has a common height of approximately $0.44958I_0$, while its impact position moves from $\beta\simeq6.59708$ at $q=-0.6$ to $\beta\simeq7.35066$ at $q=0.6$. Closer to $\beta_{\mathrm c}$, the finer rings record the additional disk crossings. Their shared critical radius and changing separation make the role of the radial geometry visible within the same emitting configuration.

\begin{figure}[h]
\centering
\includegraphics[width=0.5\linewidth]{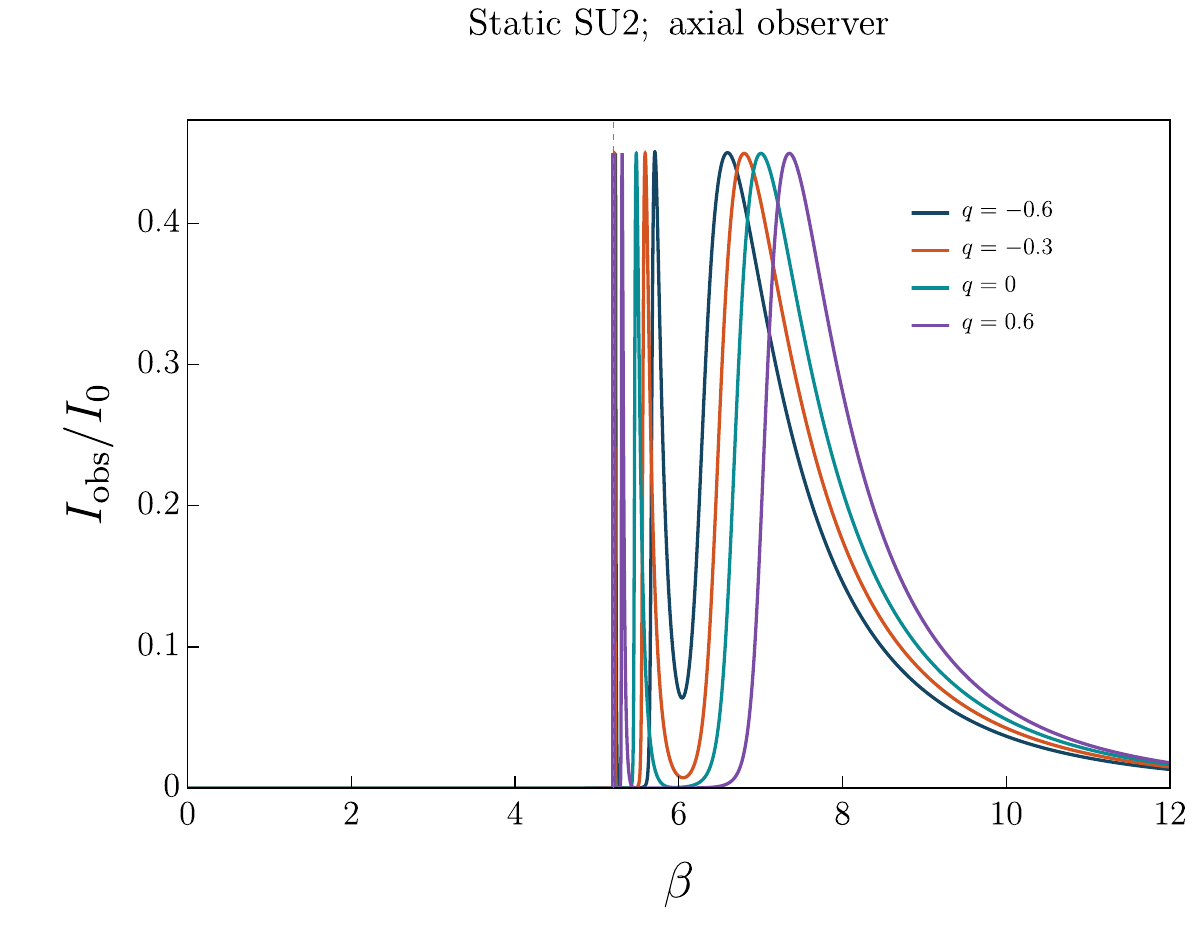}
\caption{Axial bolometric intensity for the same static SU2 source and the four couplings in Fig.~\ref{fig:axialmaps}. The curves show the broad direct contribution and the narrow features associated with additional crossings, at a common emitted scale $I_0$.}
\label{fig:axialstatic}
\end{figure}

\begin{figure}[h]
\centering
\includegraphics[width=0.43\linewidth]{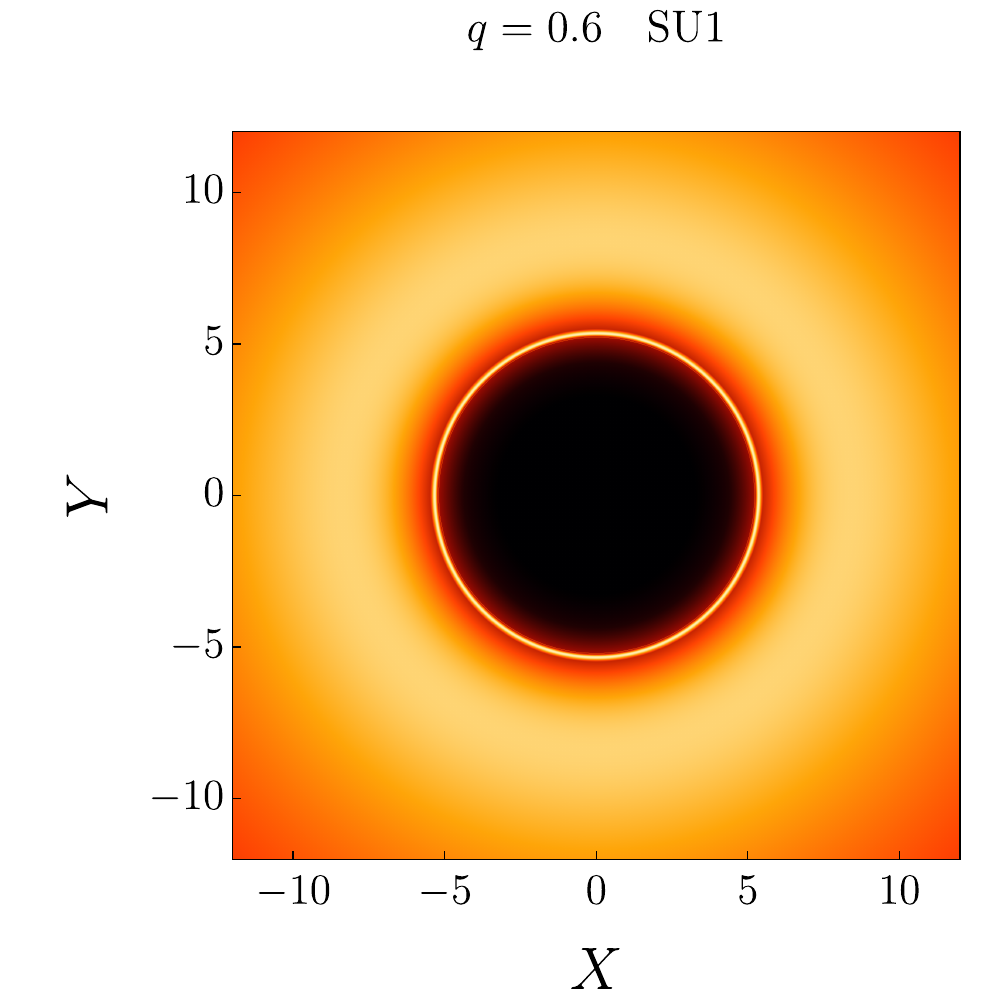}\hfill\includegraphics[width=0.43\linewidth]{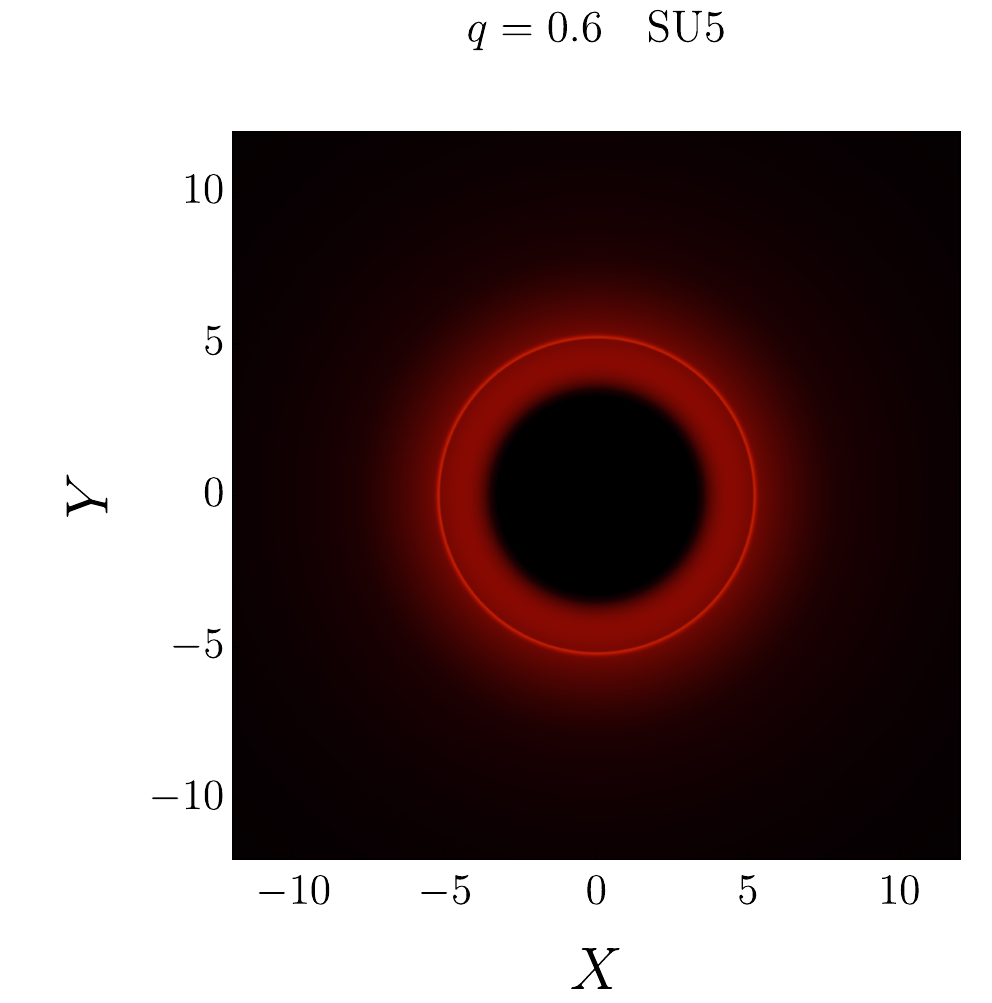}\hfill\includegraphics[height=0.32\linewidth]{color_scale.pdf}
\caption{Source dependence of the axial image at $q=0.6$. The static SU1 and SU5 sheets use the parameters specified below Eq.~\eqref{eq:su} and share the same peak emitted intensity $I_0$. The panels have one observed intensity reference, common to this pair.}
\label{fig:sources}
\end{figure}

The source dependence can be examined at fixed geometry in Fig.~\ref{fig:sources}. The SU1 and SU5 prescriptions assign emission to different radial regions, with the same peak emitted intensity $I_0$. The broader SU1 distribution illuminates a larger area of the screen, while the central concentration of SU5 gives greater weight to radiation emitted at small radius. Gravitational attenuation and the transfer functions then determine how this emission reaches the observer. The common color scale makes their brightness contrast explicit, alongside the unchanged capture threshold \cite{Rosa2022,Olmo2023,Olmo2025}.

\begin{figure}[h]
\centering
\includegraphics[width=0.33\linewidth]{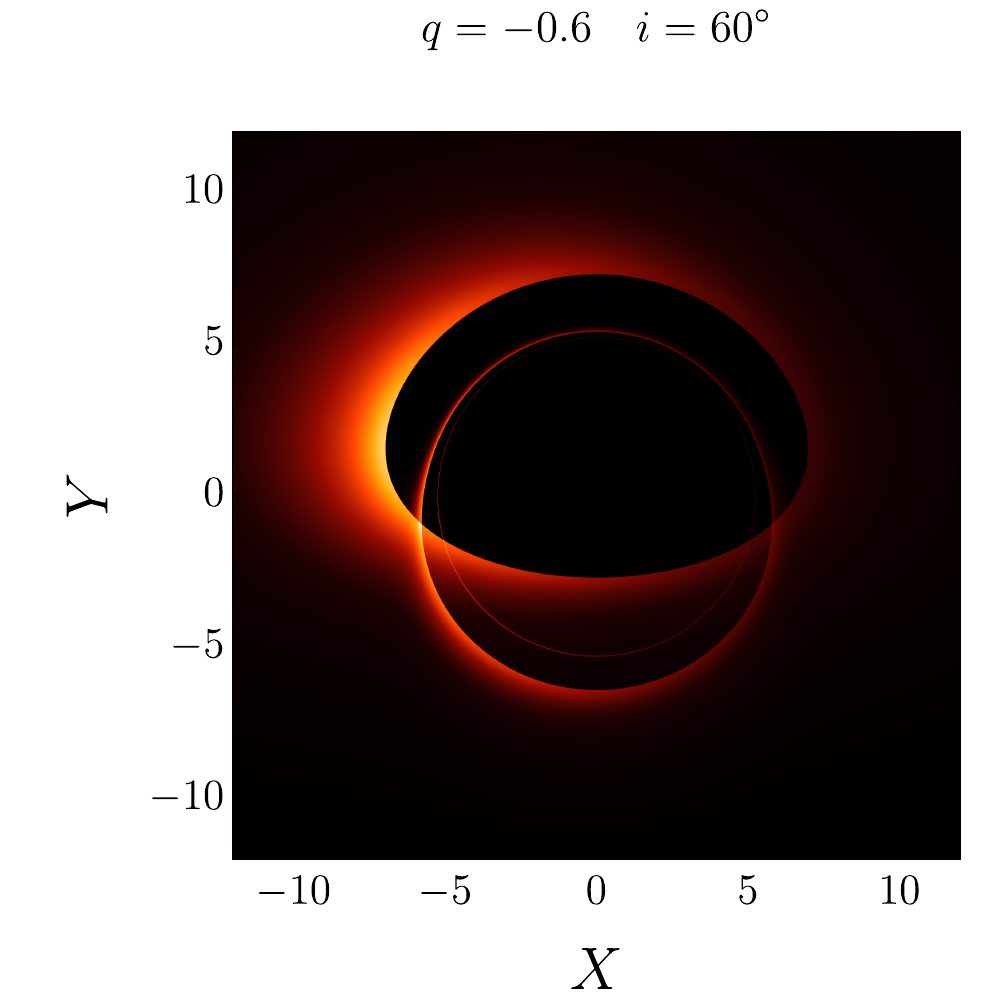}\includegraphics[width=0.33\linewidth]{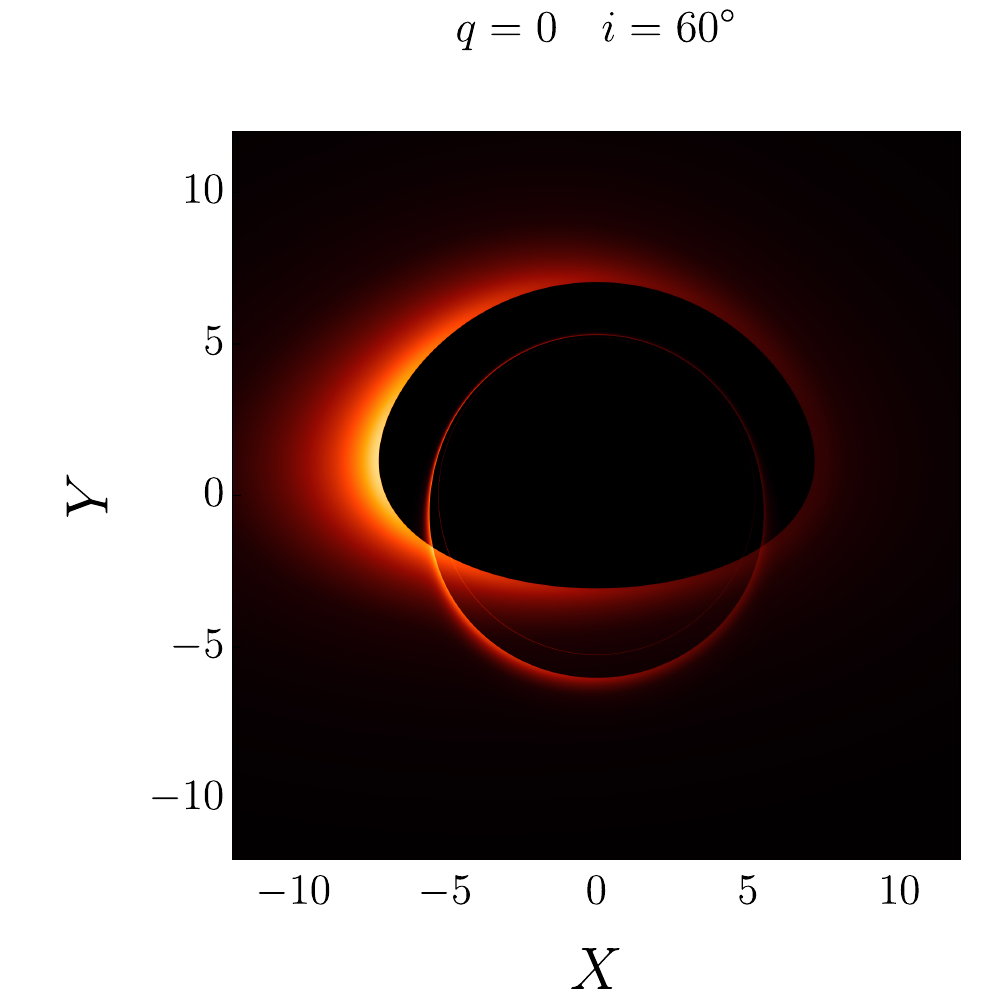}\includegraphics[width=0.33\linewidth]{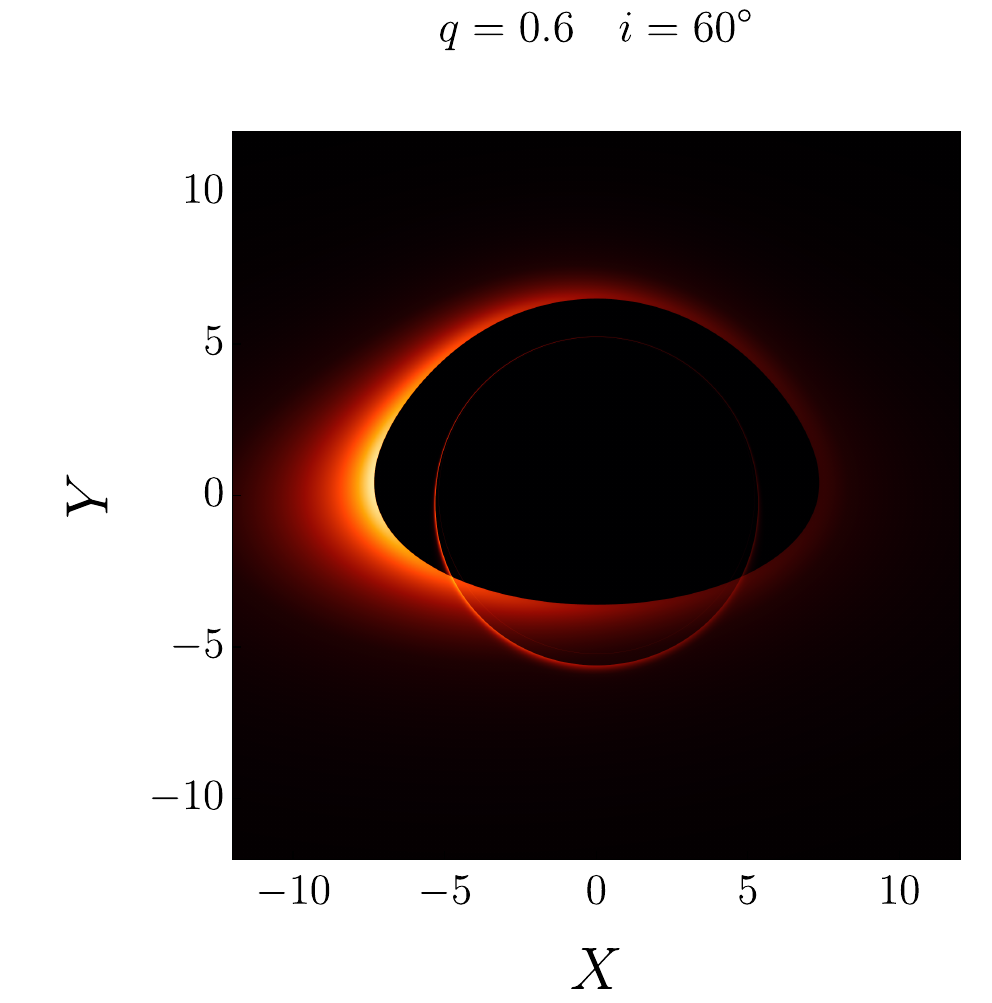}\includegraphics[height=0.27\linewidth]{color_scale.pdf}
\caption{Inclined images of the transparent SU2 sheet with circular emitters, at $i=60^\circ$ and $q=-0.6,0,0.6$ from left to right. All panels share the orbital mass, radial emission profile, field of view, and intensity scale. }
\label{fig:inclined}
\end{figure}

Figures~\ref{fig:intensity} and \ref{fig:inclined} place the same source in geometries with different radial deformation. The common critical curve shows a reference around which the finer structures reorganize. For positive $q$, $\sigma$ decreases and successive narrow images approach that curve more rapidly. For $q<0$, their larger asymptotic spacing offers a wider interval in which emission from distinct crossings can appear. The broad direct contribution samples the radial profile over a larger screen interval, with its shape set jointly by the source and the transfer function. This behavior makes the image hierarchy useful even when the characteristic capture scale is shared, in the sense emphasized in studies of photon ring structure \cite{Gralla2019,Johnson2020,Olmo2023}.

The inclined circular images additionally show the Doppler enhancement on the approaching side. Its strength at each source radius follows Eq.~\eqref{eq:shift}, while the visible arrangement of the approaching and receding emission follows the ray mapping. Likewise, a central brightness depression reflects the emitting support and frequency shifts as well as capture. Its visible edge can differ from the critical curve, as demonstrated for bosonic stars and other compact geometries \cite{Rosa2022,Rosa2023,Olmo2025}. Here, this distinction is particularly transparent: $b_{\mathrm c}/\mu$ is fixed exactly, and the radiative pattern responds to the radial geometry and the selected source.

\begin{figure}[h]
\centering
\includegraphics[width=0.33\linewidth]{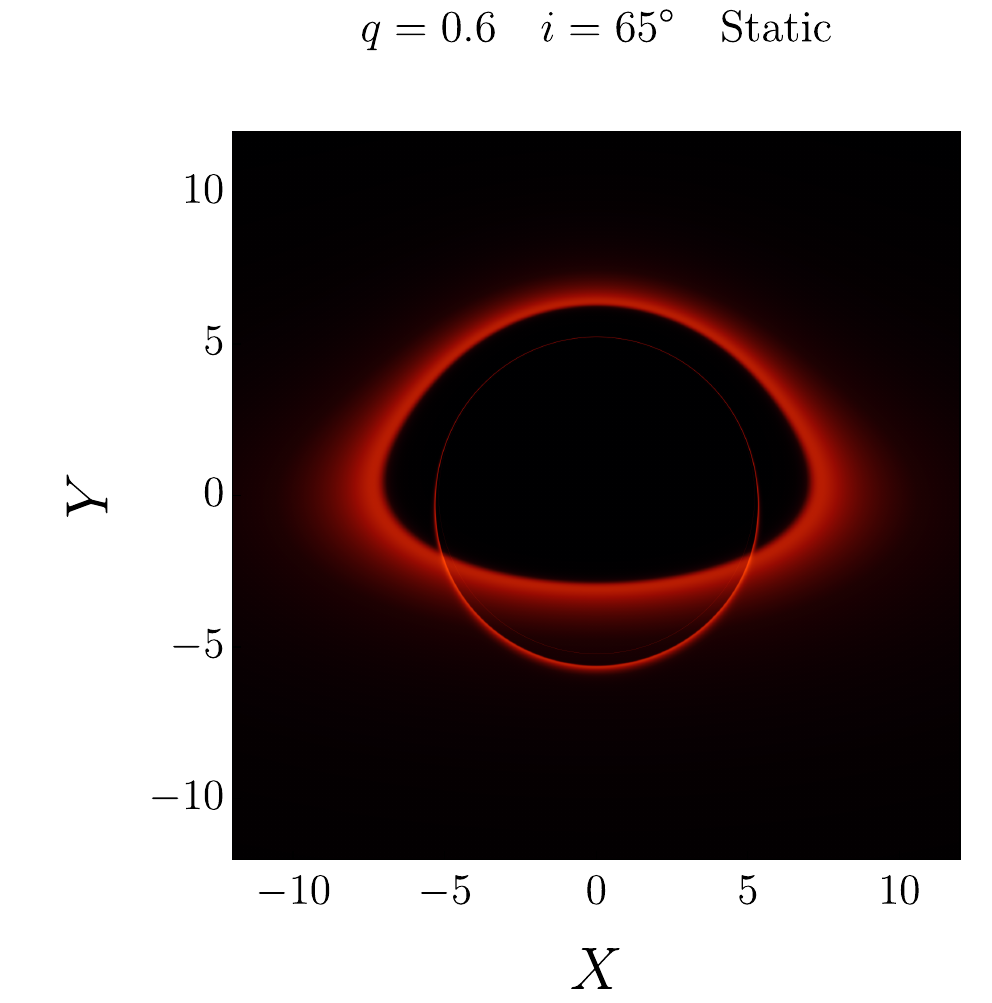}\includegraphics[width=0.33\linewidth]{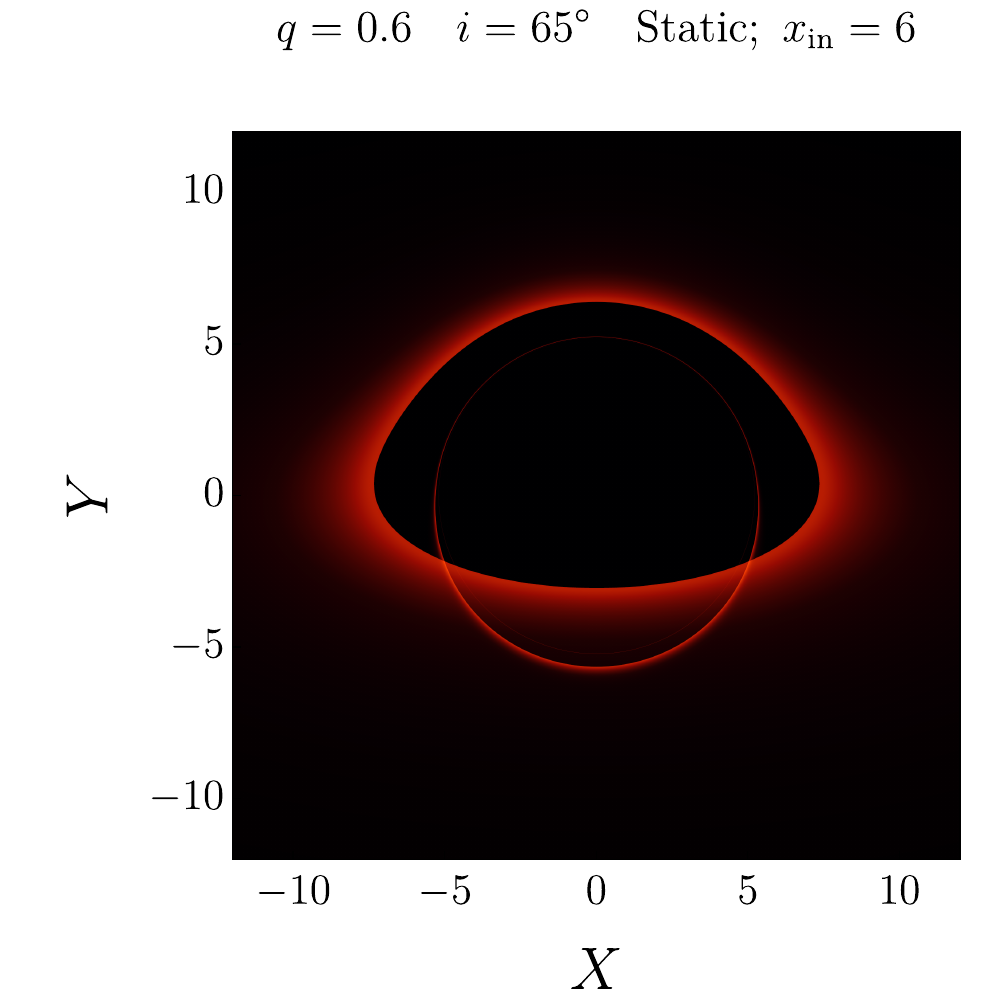}\includegraphics[width=0.33\linewidth]{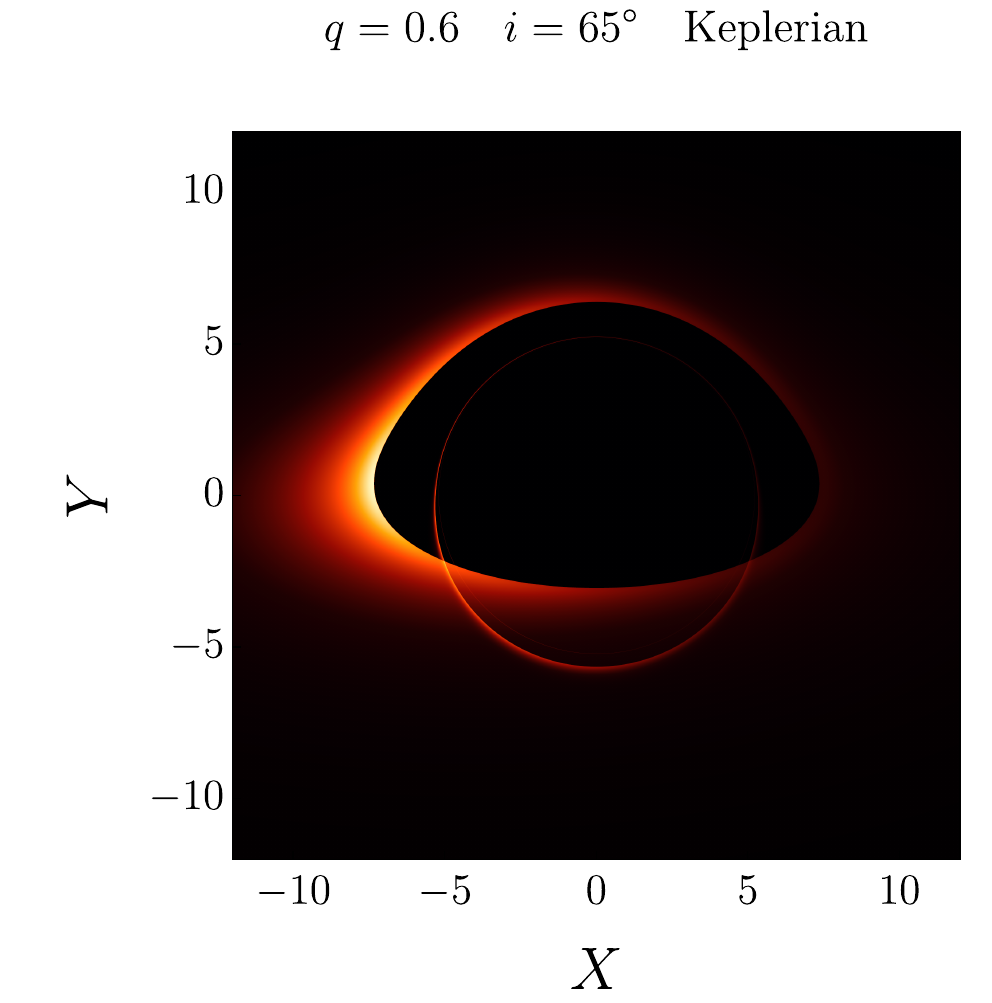}\includegraphics[height=0.27\linewidth]{color_scale.pdf}
\caption{Inclined emission at $i=65^\circ$ and $q=0.6$. From left to right: supported static emitters with $x_{\mathrm{in}}=2$, supported static emitters with $x_{\mathrm{in}}=6$, and circular emitters with $x_{\mathrm{in}}=6$. The profile parameters, emitted normalization, camera, and observed intensity reference are common to all three images.}
\label{fig:inclinedmotion}
\end{figure}

Figure~\ref{fig:inclinedmotion} examines the motion of the source at $i=65^\circ$ and $q=0.6$. The first static sheet extends from the limiting horizon edge; the second static sheet begins at $6\mu$, matching the emitting support of the circular source in the third panel. This intermediate image identifies the contribution of the inner cutoff. Orbital motion introduces the approaching and receding contrast through Eq.~\eqref{eq:shift}, while every panel samples the same intrinsic SU2 profile. In this regard, the image comparison takes into account both the radial extent of the source and its state of motion \cite{Luminet1979,Cunningham1975,Olmo2025}.

\section{Coupling sensitivity and numerical analysis}
\label{sec:assessment}

The analytic relations above separate quantities controlled by circular binding energy from those that sample proper radial distance. Their small coupling expansions are
\begin{equation}
\begin{gathered}
\frac{\mathcal F_q}{\mathcal F_0}=1+\frac q4+O(q^2),\qquad \frac{T_q}{T_0}=1+\frac q{16}+O(q^2),\qquad \frac{n_q}{n_0}=1-\frac q4+O(q^2),\\ \ln\frac{\Delta\beta_{m+1}}{\Delta\beta_m}=-\pi\left[1+\frac q4+O(q^2)\right].
\end{gathered}
\label{eq:sensitivity}
\end{equation}
The subscripts $q$ and $0$ refer to the deformed geometry and Schwarzschild at the same $\mu$, $\dot M_0$, and source parameters. The ratios involving flux and temperature apply at $x>6$. A measurement of the common critical radius sets the orbital scale; additional information about the radial image hierarchy and the matter supply would be needed to extract $q$. Geometric observations in this vacuum sector determine the combination $\ell_2/(1-\ell_1)$, leaving the individual curvature couplings degenerate \cite{Pereira2026Fingerprints}.

\begin{table}[h]
\centering
\caption{Exact rescaling factors at fixed orbital mass and rest mass supply. The final column gives the asymptotic separation ratio for successive disk images.}
\label{tab:scalings}
\begin{tabular}{rrrrr}
\toprule
$q$ & $\sigma$ & $\mathcal F_q/\mathcal F_0$ & $T_q/T_0$ & $\exp(-\pi/\sigma)$ \\
\midrule
$-0.6$ & $1.140175$ & $0.877058$ & $0.967736$ & $0.063586$ \\
$0$ & $1$ & $1$ & $1$ & $0.043214$ \\
$0.6$ & $0.836660$ & $1.195229$ & $1.045593$ & $0.023402$ \\
\bottomrule
\end{tabular}
\end{table}

The values in Table~\ref{tab:scalings} and the figures illustrate the mechanism over an enlarged coupling interval. Under the universal coupling and Solar System assumptions adopted in Ref.~\cite{She2026PalatiniKR}, the reported perihelion bound is $-7.4\times10^{-12}\leq q\leq3.7\times10^{-11}$. Equation~\eqref{eq:sensitivity} gives fractional changes of order $10^{-11}$ in the local flux and order $10^{-12}$ in temperature.

The numerical construction uses the Schwarzschild photon equation and the exact angular rescaling. Horizon and escape events locate the physical angular interval, on which the equation is integrated to obtain smooth interpolation. The conserved first integral in Eq.~\eqref{eq:null}, a direct integration of the deformed equation, and an independent radial quadrature supply a natural check on the propagation. Close to $\beta_{\mathrm c}$, Eq.~\eqref{eq:elliptic} supplies the refined radial intensity samples and is checked against independent integration at increased precision. The thermal flux is evaluated both from Eq.~\eqref{eq:flux} and its analytic expression, and its energy integral is checked against Eq.~\eqref{eq:luminosity}. On $32$ sampled production rays, the maximum dimensionless residual $|\beta^2[\sigma^2(\mathrm{d}u/\mathrm{d}\chi)^2+u^2-2u^3]-1|$ is below $1.3\times10^{-11}$. Direct quadrature of the accretion integral agrees with Eq.~\eqref{eq:analyticflux} to a relative error below $9\times10^{-15}$ at the checked radii, and the integrated efficiency agrees with Eq.~\eqref{eq:luminosity} within $7\times10^{-16}$.

The inclined coupling comparisons use $401\times401$ pixels at $i=60^\circ$, and the source motion comparison uses $501\times501$ pixels at $i=65^\circ$. Their common field of view is $-12\leq X,Y\leq12$. The face on maps contain $1001\times1001$ pixels, with a $2\times2$ subpixel average. Their intensity is interpolated separately on the two sides of $\beta_{\mathrm c}$, using $15751$ radial samples and independent midpoint checks. Outside $|\beta-\beta_{\mathrm c}|<0.012$, the sampled interpolation errors are below $2\times10^{-4}$ of the profile peak. The narrower critical strip is evaluated with the elliptic solution and examined through pixel averaging. On a representative $501\times501$ map, increasing the subpixel average from $2\times2$ to $4\times4$ changes the screen integrated intensity by $9.57\times10^{-5}$ in relative terms. The axial profiles of circular emitters use $2802$ impact samples, combining a regular spacing of $0.012$ with logarithmic refinement around $\beta_{\mathrm c}$. Each comparison shares its source normalization and intensity stretch. The raster maps show the combined emission at their stated resolution, while the sampled radial transfer functions describe structures finer than an image pixel. Notice that higher image orders require progressively smaller impact intervals according to Eq.~\eqref{eq:subrings}; their visibility also depends on optical depth, angular resolution, and the radial emissivity \cite{Johnson2020,Olmo2023,Olmo2025}. 


\section{Conclusion}

We examined the thermal emission and optical appearance of accretion disks around a static black hole in Palatini Kalb Ramond gravity. At fixed orbital mass, the circular orbital quantities and the characteristic radii shared their Schwarzschild values, while local dissipation and photon propagation responded to the radial deformation. In this sense, the common orbital scales gave a reference for addressing the coupling through disk emission and the structure of successive images.

The accretion calculation yielded an elementary expression for the emitted flux, whose amplitude scaled as $1/\sigma$. The effective temperature followed the corresponding fourth root, and the hottest annulus lay at $r\simeq9.55\mu$. Taking into account the proper radial measure gave an emitted efficiency of approximately $5.72\%$. For positive $q$, the increased local emission reflected a redistribution of the available orbital binding energy over the proper disk surface.

Moreover, the exact angular rescaling of photon trajectories determined the lensing structure. The strong deflection coefficient equaled $\sigma$, and the displacements of successive disk images from the critical curve followed the asymptotic ratio $\exp(-\pi/\sigma)$. As $q$ increased, these images accumulated more rapidly around the common capture scale. The thermal and transparent source models further showed how radial emissivity, optical depth, inclination, and orbital motion shaped the observed brightness.

Under the universal coupling assumptions associated with the quoted Solar System bounds, the fractional corrections were of order $10^{-11}$ in flux and $10^{-12}$ in temperature. The larger couplings displayed in the images illustrated the geometric mechanism over an expanded parameter interval. Within the vacuum sector considered here, the geometric observables depended on the combination $q=\ell_2/(1-\ell_1)$.


\section*{Acknowledgments}
\hspace{0.5cm} A. A. Araújo Filho is supported by Conselho Nacional de Desenvolvimento Cient\'{\i}fico e Tecnol\'{o}gico (CNPq) -- [150223/2025-0].

\section*{Data Availability Statement}

Data associated with this study consist of the analytical expressions and numerical figures presented in the manuscript. No additional dataset is required to reproduce the analytical results. The code used to generate the numerical results and figures is available from the corresponding author upon reasonable request.

\bibliography{main}

\end{document}